\documentclass[a4paper,11pt]{article}
\pdfoutput=1

\usepackage{jheppub}

\usepackage[usenames,dvipsnames]{xcolor}
\usepackage[english]{babel}
\usepackage{enumerate,enumitem,float,graphicx,comment}

\usepackage[T1]{fontenc}
\usepackage[utf8]{inputenc}
\usepackage{lmodern}
\usepackage[normalem]{ulem}

\makeatletter
\g@addto@macro\bfseries{\boldmath}
\makeatother

\usepackage{amsmath,amssymb,mathtools}

\usepackage{tabulary}
\newcolumntype{K}[1]{>{\centering\arraybackslash}p{#1}}

\newcommand{\red}{\color{red}}
\newcommand{\blue}{\color{blue}}

\newcommand{\eq}[1]{Eq.(\ref{#1})}

\def\be{\begin{equation}}
\def\ee{\end{equation}}
\def\ba{\begin{eqnarray}}
\def\ea{\end{eqnarray}}

\def\td{\tilde}
\def\pp{\partial}
\def\dagg{\dagger}
\def\mm{\mathrm}
\def\mc{\mathcal}

\def\blue{\color{blue}}

\def\blue{\color{black}}
\def\red{\color{black}}

\title{%
 Weyl double copy and massless free-fields in curved spacetimes}

\author{Shanzhong Han}
\emailAdd{shanzhong.han@nbi.ku.dk}

\affiliation{%
  The Niels Bohr Institute,
  University of Copenhagen,
  \\
  Blegdamsvej 17,
  DK-2100 Copenhagen Ø,
  Denmark
}

\abstract{%\%
{\red In spinor formalism,} since any massless free-field {\red spinor} with spin higher than $1/2$ can be constructed {\red with spin-1/2 spinors (Dirac-Weyl spinors) and scalars}, we introduce a map between {\red Weyl} fields and {\red Dirac-Weyl} fields. {\red We determine the corresponding Dirac-Weyl spinors in a given empty spacetime.} Regarding them as basic units, other higher spin massless free-field spinors are then {\red identified}. {\red Along} this way, {\red we find} some hidden fundamental features related to {\red these} fields. In particular, for {\red non-twisting} vacuum {\red Petrov} type N {\red solutions}, we show that all higher spin massless free-field {\red spinors} can be constructed {\blue with} {\red one} type of Dirac-Weyl spinor and {\red the zeroth copy}. Furthermore, we systematically rebuild the Weyl double copy for non-twisting vacuum type N and vacuum type D solutions. {\red Moreover, w}e show that the zeroth copy not only connects the gravity fields with a single copy but also {\red connects the} degenerate {\red Maxwell} fields with the Dirac-Weyl fields in the curved spacetime, both for type N and type D cases. {\red Besides}, we extend the study to non-twisting vacuum type III {\red solutions.  W}e find a {\red particular} Dirac-Weyl scalar independent of the proposed map {\blue and} whose square {\red is} proportional to the Weyl scalar. A degenerate Maxwell field and an auxiliary scalar field are then {\red identified}. Both of them play similar roles as the {\red Weyl} double copy. The result further {\red inspires us} that there {\red is} a deep connection between gravity theory and gauge theory. }

\begin{document}
\maketitle
\flushbottom

\section{Introduction}%
\label{sec:intro}
In recent years, the attempt to look for the connection between gravity theory and quantum theory has been actively investigated. {\red As is known,} Yang-Mills gauge theory {\red is by far the most successful theory to describe} the micro world{\red. On the other hand,}  given the experimental confirmation of gravitational waves \cite{LIGOScientific:2014pky,LIGOScientific:2016wof}, Einstein's gravity theory is further confirmed as the most promising theory to describe the macro-scale universe. Therefore, it is significant to {\red explore} the relationship between these two theories. {\red Much} work has been devoted to this study. One such attempt is the double copy. It start{\red ed} from the {\red research} of perturbative scattering amplitudes \cite{KAWAI19861,Bern:2008qj,Bern:2010ue,Bern:2019prr}{\red ; with the help of exact gravity solutions,} the study {\red was extended} into {\red the} classical {\red context} \cite{Monteiro:2014cda}. There are two classes of double cop{\red ies}: Kerr-Schild double copy \cite{Monteiro:2014cda,Berman:2018hwd,Luna:2015paa,Ridgway:2015fdl,White:2016jzc,Adamo:2017nia,DeSmet:2017rve,Bahjat-Abbas:2017htu,Carrillo-Gonzalez:2017iyj,Ilderton:2018lsf,Lee:2018gxc,Gurses:2018ckx,Elor:2020nqe} and Weyl double copy \cite{Luna:2018dpt,Keeler:2020rcv,Godazgar:2020zbv,White:2020sfn,Chacon:2021wbr,Chacon:2021hfe,Adamo:2021dfg,Godazgar:2021iae,Easson:2021asd}. {\red T}he latter {\red covers} a {broader} range of spacetimes so we will focus on the Weyl double copy in this paper.

In spinor formalism, the Weyl double copy is written as
\be\label{sec1:WDC}
\Psi_{ABCD}=\frac{\Phi_{(AB}\Phi_{CD)}}{S},
\ee
which maps {\red a} Weyl spinor {\red $\Psi_{ABCD}$} ({\red a} vacuum gravity fiel{\red d}) into {\red a} single copy {\red $\Phi_{AB}$} ({\red a} Maxwell field which satisfies {\red the} Maxwell equation {\red in} Minkowski spacetime) and a zeroth copy {\red $S$} ({\red a} scalar field which satisf{\red ies} the wave equation {\red in} Minkowski spacetime). 

{\red D}ecades ago, some works \cite{Walker:1970un,Hughston:1972qf}  {\red had already} given us the prediction {\red about the Weyl double copy}. In Ref. \cite{Walker:1970un}, {\blue given} a Weyl spinor of a vacuum type D solution {\red on a dyad $(o_A, \iota_A)$}, $\Psi_{ABCD}$ {\red $=\psi o_{(A}o_B\iota_C\iota_{D})$}, Walker and Penrose showed that there exists a Killing spinor of valence two $\chi_{BC}={\red \psi}^{-1/3}o_{(B}\iota_{C)}$, which satisfies {\red the} twistor equation $\nabla_{(A}^{A'}\chi_{BC)}=0$. Based on this, the same authors{\red ,} together with Hughston and Sommers{\red ,} proposed \cite{Hughston:1972qf} that in any vacuum type D spacetime with a Weyl spinor $\Psi_{ABCD}$, one can construct a test electromagnetic {\red field}, such that $\Phi_{AB}=\psi^{2/3}o_{(A} \iota_{B)}$\footnote{It might be more enlightening if we write it in form of \eq{sec1:WDC}, the difference is the background that {\red the} Maxwell and scalar field{\red s} are living in is a curved spacetime instead of a Minkowski spacetime.}. This work {\red discovered an intriguing} relation between gravity and Maxwell fields in curved spacetimes. {\red Combined} with the fact that {\red the} Maxwell field is the simplest solution of the gauge theory {\red ---} the case of the group $U(1)$,  the Weyl double copy {\red relation appears to be more essential} between gravity theory and gauge theory. It was proposed for the first time {\red for} vacuum type D {\red solutions} \cite{Luna:2018dpt}. Then the Weyl double copy was proven to work {\red also} for non-twisting vacuum type N solutions \cite{Godazgar:2020zbv}. For {\red the} type III case, {\red using} the twistor theory, {\red the} study only showed it holds at the linearised level \cite{White:2020sfn,Chacon:2021wbr}. On the other hand, the asymptotic behaviours of {\red the Weyl} double copy have been discussed in recent works \cite{Adamo:2021dfg,Godazgar:2021iae}, which {\red state} that the Weyl double copy holds asymptotically for the algebraically general solutions by using the {\blue peeling} property of the Weyl scalars \cite{Newman:1961qr,Wald:1984rg}.  More recently, Ref. \cite{Easson:2021asd} studied the Weyl double copy for general type D spacetimes with external sources but without a cosmological constant and introduced an extended Weyl double copy. However{\red ,} Weyl double copy for a general spacetime, even for vacuum spacetime with a cosmological constant, is still unknown. More generally speaking, our understanding of the connection between gravity theory and gauge theory {\red remains to} be improved. Hopefully, there are {\red many exciting} and promising roads {\red a}waiting us. In particular, although the Weyl double copy prescription does not capture the current double copy interpretation of twisting type N solution{\red ,} which might require a more general and complicated prescription, {\red the} curved double copy {\red indeed} holds in this case. {\red This fact leads us to consider that it might be helpful to study first the map between a gravity field and a test Maxwell field in the curved spacetime. Or, it might be worth exploring first} the features of spin-$n/2$ $(n=0,1,2,3)$ massless free fields {\red that live} in the curved spacetime. Then the curvature information {\red will} be reflected by these lower spin fields. By probing the features of these fields, it would be easier to look for those curvature{\red -}independent fields, such as pure Maxwell field$\footnote{Where "pure Maxwell fields" means {\red that the} Maxwell fields are living in Minkowski spacetime as the special solutions of gauge theory, so they are totally independent of the gravity theory.}$.  In this paper, regarding spin-$1/2$ massless free-field spinors{\red ---} Dirac-Weyl (DW) spinors{\red ,} as basic units, we not only identify {\red the} DW spinor{\red s} but also construct higher massless free-field spinor{\red s} following the proposed map. Then, one will see that the relations between gravity fields and Maxwell fields in curved spacetime found in \cite{Hughston:1972qf,Godazgar:2020zbv} are a {\red particular} case {\red in the present work;} more fundamental properties {\red of} these {\red fields} are revealed. Especially, a natural map similar to {\red the Weyl} double copy from gravity fields to pure Maxwell fields in type III spacetime is proposed with the aid of {\red a} scalar field.

The structure of our paper is as follows. In section \ref{sec:Massless free-field in spinor formalism}, we give a brief review of spinor algebra and the massless free-field spinors. Section \ref{sec3:from gravity to lower spin massless free fields} identif{\red ies} the DW spinors {\red in vacuum type N, type D, and type III spacetimes, respectively}. Regarding them as basic units, we anali{\red z}e the properties of different spin massless free-field{\red s}, especially {\red the DW} and Maxwell {\red field}s. Then we systematically reconstruct the Weyl double copy for non-twisting vacuum type N and vacuum type D spacetimes. A new property of the zeroth copy is discovered after that.  Following this {\red way}, a degenerate electromagnetic field {\red that lives} in Minkowski spacetime and an associated scalar field are obtained from the vacuum type III solutions. The discussion and conclusions are given in section \ref{sec4:discussion and conclusion}.

%%%%%%%%%%%%%%%%%%%%%%%
\section{Massless free-fields in spinor formalism}
\label{sec:Massless free-field in spinor formalism}

Since massless free-field equations have a simple form in spinor formalism, before going on, we shall give a short introduction {\red to} spinor algebra{\red ;} for more details, one may refer to Refs. \cite{1987ssv..book.....P,1994agr..book.....S}.

First, let us consider an arbitrary vector {\red V on} the basis ${e_{i}}$,
\be
V=V^0e_0+V^1e_1+V^2e_2+V^3e_3{\red .}
\ee
{\red W}e transfer it to a $2\times2$ Hermitian matrix
\be\label{sec2:from real to complex}
H=V^a\sigma_a=V^0\sigma_0+V^j\sigma_j=\begin{pmatrix} 
                                                                  V^0+V^3 & V^1+iV^{\red 2} \\ 
                                                                  V^1-i V^2& V^0-V^3 
                                                                  \end{pmatrix},
\ee
where $\sigma_0$ is a $2\times2$ unit matrix, {\red and} $\sigma_j\ (j=1,2,3)$ are Pauli spin matrices. By introducing a pair of complex numbers {\red $(\xi,\eta)$} as {\red follows,}
\be
\begin{array}{ll}
V^0=\frac{1}{\sqrt{2}}(\xi \bar{\xi}+\eta \bar{\eta}), & V^1=\frac{1}{\sqrt{2}}({\red \xi} \bar{\eta}+\eta {\red \bar{\xi}}), \\
V^2=\frac{1}{i \sqrt{2}}(\xi \bar{\eta}-\eta {\red \bar{\xi}}), & V^3=\frac{1}{\sqrt{2}}(\xi {\red \bar{\xi}}-\eta \bar{\eta}),
\end{array}
\ee
where the bar denotes the operation of complex conjugation, \eq{sec2:from real to complex} {\red can be translated into a new form}
\be\label{sec2: spin transformation}
\frac{1}{\sqrt{2}}\begin{pmatrix} 
V^0+V^3 & V^1+iV^{\red 2} \\ 
V^1-i V^2& V^0-V^3 
\end{pmatrix}
=
\left(\begin{array}{ll}
\xi \bar{\xi} & \xi \bar{\eta} \\
\eta \bar{\xi} & \eta \bar{\eta}
\end{array}\right)
=
\left(\begin{array}{l}
\xi \\
\eta
\end{array}\right)\left(\begin{array}{ll}
\bar{\xi} & \bar{\eta}
\end{array}\right).
\ee
On the other hand, there {\red is} a complex linear transformation of pair $(\xi,\eta)^T$:
\be
\left(\begin{array}{l}
\hat{\xi} \\
\hat{\eta}
\end{array}\right)
=
\left(\begin{array}{ll}
\alpha & \beta \\
\gamma & \delta
\end{array}\right)
\left(\begin{array}{l}
\xi \\
\eta
\end{array}\right)
=
A\left(\begin{array}{l}
\xi \\
\eta
\end{array}\right),
\ee
where the hat denotes the new quantity after the transformation. If we impose the condition $\operatorname{det} A=1$, it {\red corresponds to the} spin transformation{\red,} all of {\red such} matrices form the group $S L (2, \mathbb{C})$. When a spin transformation is applied, \eq{sec2: spin transformation} becomes
\be
\begin{pmatrix} 
V^0+V^3 & V^1+iV^2 \\ 
V^1-i V^2& V^0-V^3 
\end{pmatrix}
\to
A\begin{pmatrix} 
V^0+V^3 & V^1+iV^2 \\ 
V^1-i V^2& V^0-V^3 
\end{pmatrix}A^\dagg
=
\begin{pmatrix} 
\hat{V}^0+\hat{V}^3 & \hat{V}^1+i\hat{V}^2 \\ 
\hat{V}^1-i \hat{V}^2& \hat{V}^0-\hat{V}^3 
\end{pmatrix}=\hat{H},
\ee
where the dagger denotes the operation of conjugate transpose. It is worth {\red noting} that due to the condition $\operatorname{det} A=1$, the determinant of the Hermitian matrix remains invariant, 
\be
\operatorname{det} \hat{H}=\operatorname{det}H=(V^0)^2-(V^1)^2-(V^2)^2-(V^3)^2{\red .}
\ee
{\red In other words,} the norm of vector $V$ is invariant under the transformation. Therefore every matrix element $A$ of the group $S L (2, \mathbb{C})$ defines a restricted Lorentz transformation. {\red As is known}, $S L (2, \mathbb{C})$ is homomorphic to Lorentz group\footnote{Note $SO(3,1)$ is not isomorphic to $S L (2, \mathbb{C})$, not all spinors have a tensor courterpart. {\red In general}, tensors can be regarded as special cases of spinors.}. Furthermore, since $S L (2, \mathbb{C})$ is isomorphic to the symplectic group $Sp(2,\mathbb{C})$, it is natural to introduce the $2$-dimensional symplectic vector space (spin-space) over $\mathbb{C}$.  {\red A} tensor is defined in spinor form such that $T_{a...b}^{\ \ \ \ c...d}=T_{AA'...BB'}^{\ \ \ \ \ \ \ \ \ \ CC'...DD'}$ with abstract index notation. {\red In practice,} tensors commute with associated spinors  through the Infeld-van der Waerden symbols%Note, the space spinors constructing is bigger than the space tensor buiding.
\be
\Sigma_{\bold{a}}^{\bold{AA'}}=\frac{1}{\sqrt{2}}\sigma_{\bold{a}}^{(\bold{AA'})}, \quad \quad \bold{a}=0,1,2,3,\quad \bold{A}=0,1
\ee
under the component transformation relation
\be
T_{\bold{a...c}}^{\ \ \bold{d...f}}=T_{\bold{AA'...CC'}}^{\ \ \ \ \ \ \ \ \ \ \bold{DD'...FF'} }\Sigma_{\bold{a}}^{\ \bold{AA'}}...\Sigma_{\bold{{\red c}}}^{\ \bold{ CC'}}\Sigma^{\bold{d}}_{\ \bold{DD'}}...\Sigma^{\bold{f}}_{\ FF'},
\ee
where matrices $\sigma_{\bold{a}}$ are conventionally chosen as Pauli matrices, small bold latin letters denote the indices of tensor components, capital bold latin letters denote the indices of spinor components and the prime marks the indices on complex conjugate space\footnote{In spinor algebra, the complex conjugate space is anti-isomorphic with the spin-space.}, e.g. $\overline{T^{A}}=\bar{T}^{A'}$. Note that{\red , in general,} the Infeld-van der Waerden symbols are used in the specific transformation calculation between a tensor and a spinor, they will not appear in this paper. One can refer to the last part of $3.1$ of Ref. \cite{1994agr..book.....S} for more details {\red about this symbol}.

Now{\red ,} let us focus on spin-space{\red . A}ny vectors $\xi^A$ can be expanded {\red on} a {\red spinor dyad } {\red $(o,\iota)$},
\be
\xi^A=\xi^0 o^A+\xi^1 \iota^A\quad \Leftrightarrow \xi^{\bold{A}}=\left(\begin{array}{l}
\xi^0 \\
\xi^1
\end{array}\right),
\ee
where $o^A$ can be any non-zero vector, and another vector $\iota^A$ is imposed to satisfy $\{o,\iota\}=1$. Then one may find
\be
o^{\bold{A}}=\left(\begin{array}{l}
1 \\
0
\end{array}\right),\quad
\iota^{\bold{A}}=\left(\begin{array}{l}
0 \\
1
\end{array}\right).
\ee
In addition, the symplectic structure implies that the inner product of two arbitrary vectors satisfies
\be
\{\xi,\eta\}=\varepsilon_{AB}\xi^A\eta^B=-\{\eta,\xi\},
\ee
where $\varepsilon_{AB}$ plays a role analogous to the metric tensor{\red ;} nevertheless, it is anti-symmetric
\be
\varepsilon_{AB}=-\varepsilon_{BA}.
\ee
Then normalization condition reads
\be
\begin{aligned}
\{o,{\red \iota}\}=\varepsilon_{AB}o^A\iota^B=-\{\iota,o\}=-\varepsilon_{AB}\iota^Ao^B=1,\\
\{o,o\}=\varepsilon_{AB}o^Ao^B=0, \ \ \{\iota,\iota\}=\varepsilon_{AB}\iota^A\iota^B=0,
\end{aligned}
\ee
where it is easy to see
\be
\varepsilon_{\bold{AB}}=\varepsilon^{\bold{AB}}=\left(\begin{array}{ll}
0 & 1 \\
-1 & 0
\end{array}\right),
\ee
{\red and}
\be
\varepsilon_{AB}=2o_{[A}\iota_{B]},\quad \varepsilon^{AB}=2o^{[A}\iota^{B]}.
\ee
The rule of raising and lowering indices is as follows
\be
\varepsilon^{AB}\xi_B=\xi^A,\quad \xi^A \varepsilon_{AB}=\xi_B.
\ee
The relations above also hold in the complex conjugate space.

In addition, the null tetrad can be written in terms of {\red the} spinor {\red bases}
\be\label{sec2:NP}
\begin{array}{llll}
\ell^{a}=o^{A} \bar{o}^{A^{\prime}}, & n^{a}=\iota^{A} \bar{\iota}^{A^{\prime}}, & m^{a}=o^{A} \bar{\iota}^{A^{\prime}}, & \bar{m}^a=\iota^A\bar{o}^{A^{\prime}} ,\\
\ell_{a}=o_{A} \bar{o}_{A^{\prime}}, & n_{a}=\iota_{A} \bar{\iota}_{A^{\prime}}, & m_{a}=o_{A} \bar{\iota}_{A^{\prime}},&\bar{m}_a={\red \iota_A\bar{o}_{A^{\prime}}},
\end{array}
\ee
where real null vectors $\ell$ and $n$ satisfy $\ell^2=n^2=0$, $\ell \cdot n=1$, complex null vectors $m$ and $\bar{m}$ satisfy $m^2=\bar{m}^2=0$, $m\cdot \bar{m}=-1$, furthermore, $\ell\cdot m=n\cdot m=\ell\cdot \bar{m}=n\cdot \bar{m}=0$. The definitions of spin coefficients in this paper are consistent with Appendix B of Ref. \cite{1994agr..book.....S} and Ref. \cite{1998mtbh.book.....C}. They are listed as follows
\be
\begin{aligned}
&\kappa^{\red *}=m^{\bold{a}}\ell^{\bold{b}}\nabla_{\bold{b}}\ell_{\bold{a}}, \quad &\pi^{\red *}=n^{\bold{a}}\ell^{\bold{b}}\nabla_{\bold{b}}\bar{m}_{\bold{a}}, \quad &\epsilon^{\red *}=\frac{1}{2}(n^{\bold{a}}\ell^{\bold{b}}\nabla_{\bold{b}}\ell_{\bold{a}}+m^{\bold{a}}\ell^{\bold{b}}\nabla_{\bold{b}}\bar{m}_{\bold{a}}),\\
&\tau^{\red *}=m^{\bold{a}}n^{\bold{b}}\nabla_{\bold{b}}\ell_{\bold{a}}, \quad &\nu^{\red *}=n^{\bold{a}}n^{\bold{b}}\nabla_{\bold{b}}\bar{m}_{\bold{a}}, \quad &\gamma^{\red *}=\frac{1}{2}(n^{\bold{a}}n^{\bold{b}}\nabla_{\bold{b}}\ell_{\bold{a}}+m^{\bold{a}}n^{\bold{b}}\nabla_{\bold{b}}\bar{m}_{\bold{a}}),\\
&\sigma^{\red *}=m^{\bold{a}}m^{\bold{b}}\nabla_{\bold{\red{b}}}\ell_{\bold{{\red a}}},\quad &\mu^{\red *}=n^{\bold{a}}m^{\bold{b}}\nabla_{\bold{b}}\bar{m}_{\bold{a}}, \quad &\beta^{\red *}=\frac{1}{2}(n^{\bold{a}}m^{\bold{b}}\nabla_{\bold{b}}\ell_{\bold{a}}+m^{\bold{a}}m^{\bold{b}}\nabla_{\bold{b}}\bar{m}_{\bold{a}}),\\
&\rho^{\red *}=m^{\bold{a}}\bar{m}^{\bold{b}}\nabla_{\bold{b}}\ell_{\bold{a}}, \quad &\lambda^{\red *}=n^{\bold{a}}\bar{m}^{\bold{b}}\nabla_{\bold{b}}\bar{m}_{\bold{a}}, \quad &\alpha^{\red *}=\frac{1}{2}(n^{\bold{a}}\bar{m}^{\bold{b}}\nabla_{\bold{b}}\ell_{\bold{a}}+m^{\bold{a}}\bar{m}^{\bold{b}}\nabla_{\bold{b}}\bar{m}_{\bold{a}}).
\end{aligned}
\ee
{\red To distinguish from other symbols, we use the star to mark the spin coefficients.}

We have given a short introduction {\red to} spinor algebra{\red . Now}, let us turn to massless free-fields (sour{\red c}e-free). {\red We shall list the corresponding spinors without proof, one may refer to sections 5.7 of Ref. \cite{1987ssv..book.....P} for more details.} 

Given a symmetric spinor with $n$ indexes $\mc{S}_{A_1 A_2...A_n}$, spin-$n/2$ massless free-field equations are translated into a {\red simple} form
\be\label{sec2:general massless free-field equation}
\nabla^{A_1A'_1}\mc{S}_{A_1A_2...A_n}=0.
\ee

{\red When $n=4$, the spinor $\mc{S}$ refers to a} Weyl {\red spinor $\Psi_{ABCD}$ translated from the Weyl tensor}\footnote{{\red A}s an exception, $\bar{\varepsilon}_{A'B'}$ is usually abbreviated as $\varepsilon_{A'B'}$, that is also applied for raised index version. In addition, {\red one may realize} that the first {\red identity of \eq{sec2: Weyl spinor}} is a general tensor-spinor identity with abstract indices, we do not need to use the Infeld-van der Waerden symbols here.}
\be\label{sec2: Weyl spinor}
C_{abcd}=C_{AA'BB'CC'DD'}=\Psi_{ABCD}\varepsilon_{A'B'}\varepsilon_{C'D'}+\bar{\Psi}_{A'B'C'D'}\varepsilon_{AB}\varepsilon_{CD}.
\ee
{\red Following the vacuum Einstein's field equation, \eq{sec2:general massless free-field equation} in this case represents} the Bianchi identity (with or without a cosmological constant) {\red }
\be\label{sec2:vacuum Einstein equation}
\nabla^{AA'}\Psi_{ABCD}=0.
\ee 
According to Petrov classification, there are five different types of solutions:
\be
\begin{aligned}
I: \Psi_{A B C D} & \sim \td{\alpha}_{(A} \td{\beta}_{B} \td{\gamma}_{C} \td{\delta}_{D)}, \\
I I: \Psi_{A B C D} & \sim \td{\alpha}_{(A} \td{\alpha}_B \td{\gamma}_{C} \td{\delta}_{D)}, \\
I I I: \Psi_{A B C D} & \sim \td{\alpha}_{(A} \td{\alpha}_B \td{\alpha}_{C} \td{\delta}_{D)}, \\
D: \Psi_{A B C D} & \sim \td{\alpha}_{(A} \td{\alpha}_B \td{\delta}_{C} \td{\delta}_{D)}, \\
N: \Psi_{A B C D} & \sim \td{\alpha}_{(A} \td{\alpha}_B \td{\alpha}_{C} \td{\alpha}_{D)},
\end{aligned}
\ee
where $\td{\alpha}, \td{\beta}, \td{\gamma}$ and $\td{\delta}$ are four different non-proportional and non-vanishing {\red s}pinors. {\red T}he tilde is used to {\red distinguish them from} the spin coefficients. In addition, with the help of Newman-Penrose formalism, {\red the} Weyl tensor {\red is} reduced to five independent complex {\red scalars}{\red ,} 
\be
\begin{aligned}
\psi_0 &=C_{abcd}\ell^a m^b \ell^c m^d =\Psi_{ABCD}o^A o^B o^C o^D,\\
\psi_1 &=C_{abcd}\ell^a m^b \ell^c n^d =\Psi_{ABCD}o^A o^B o^C \iota^D,\\
\psi_2 &=C_{abcd}\ell^a m^b \bar{m}^c n^d =\Psi_{ABCD}o^A o^B \iota^C \iota^D,\\
\psi_3 &=C_{abcd}\ell^a n^b \bar{m}^c n^d =\Psi_{ABCD}o^A \iota^B \iota^C \iota^D,\\
\psi_4 &=C_{abcd}\bar{m}^a n^b \bar{m}^c n^d =\Psi_{ABCD}\iota^A \iota^B \iota^C \iota^D.
\end{aligned}
\ee
The second set of equalities {\red is} obtained from \eq{sec2:NP} and \eq{sec2: Weyl spinor}. Then {\red the} Weyl spinor can be {\red expanded} in a general form
\be\label{sec2:general Weyl}
\begin{aligned}
\Psi_{ABCD} &=\psi_0 \iota_A\iota_B\iota_C\iota_D-4\psi_1o_{(A}\iota_B\iota_C\iota_{D)}+6\psi_2o_{(A}o_B\iota_C\iota_{D)}\\
&-4\psi_3o_{(A}o_Bo_C\iota_{D)}+\psi_4o_{A}o_Bo_C o_{D}.
\end{aligned}
\ee

%We will consider spin 3/2 massless fields in more details in coming days.
{\red When $n=2$, the spinor $\mc{S}$ refers to an} electromagnetic {\red spinor $\Phi_{AB}$ translated from the Maxwell tensor}
\be\label{sec2:MaxwelltensorTospinor}
F_{ab}=F_{AA'BB'}=\Phi_{AB}\varepsilon_{A'B'}+\bar{\Phi}_{A'B'}\varepsilon_{AB}.
\ee
{\red \eq{sec2:general massless free-field equation} in this case represents t}he source-free Maxwell equation
\be\label{sec2:spin1eom}
\nabla^{AA'}\Phi_{AB}=0.
\ee
In analogy to {\red the} Weyl spinor, there are two different types {\red of Maxwell spinors}: 
\be\label{sec2:general Maxwell spinors}
\begin{aligned}
I: \Phi_{AB} & \sim \td{\alpha}_{(A}\td{\delta}_{B)}, \\
N: \Phi_{AB} & \sim \td{\alpha}_A \td{\alpha}_B,
\end{aligned}
\ee
where $\td{\alpha}_A$ and $\td{\delta}_A$ are two non-proportional spinors. {\red We also call Type N Maxwell spinor as degenerate Maxwell spinor. Because the corresponding {\blue electric} fields $\bold{E}$ and magnetic fields $\bold{B} $ are of the same magnitude and they are perpendicular; namely, $|\mathbf{B}|^{2}-|\mathbf{E}|^{2}=0$, $\quad \mathbf{B}\cdot\mathbf{E}=0$. In addition, for later convenience, we define three typical Maxwell spinors as follows,
\begin{align}
\text{Type I:\ \ \ \ \ \ \ }&\notag\\
\Phi^{(1)}_{AB}&=\phi_1o_{(A}\iota_{B)}\label{sec2:typeIMaxwell spinor},\\
\text{Type N:\ \ \ \ \ \ \ }&\notag\\
\Phi^{(0)}_{AB}&=\phi_0\iota_A\iota_B\label{sec2:typeN0Maxwell spinor},\\
\Phi^{(2)}_{AB}&=\phi_2o_Ao_B\label{sec2:typeN2Maxwell spinor},
\end{align} 
where the coefficients $\phi_1, \phi_0$ and $\phi_2$ are called Maxwell scalars. They are expanded in three different ways in the spin space. Substituting \eq{sec2:typeIMaxwell spinor} into \eq{sec2:spin1eom}, then multiplying $o^B$ and $\iota^B$ on \eq{sec2:spin1eom}, respectively, we obtain two dyad components of the field equation,
\begin{align}
o_A \nabla^{AA'}\log \phi_1 - 2\iota_A o^B \nabla^{AA'}o_B=0 \label{sec2:non-degenerate Maxwell 1},\\
\iota_A \nabla^{AA'}\log \phi_1 + 2o_A \iota^B \nabla^{AA'}\iota_B=0 \label{sec2:non-degenerate Maxwell 2}.
\end{align}
Analogously, from \eq{sec2:typeN0Maxwell spinor} and \eq{sec2:typeN2Maxwell spinor} we arrive at
\begin{align}
\iota_A\nabla^{AA'}\log {\phi_0}-2\iota_A o^B\nabla^{AA'}\iota_B+o_A\iota^B\nabla^{AA'}\iota_B=0 \label{sec2:degenerate Maxwell0},\\
o_A\nabla^{AA'}\log\ \phi_2+2o_A \iota^B \nabla^{AA'}o_B-\iota_A o^B \nabla^{AA'}o_B=0 \label{sec2:degenerate Maxwell2}.
\end{align}
}
{\red Recalling \eq{sec2:MaxwelltensorTospinor}, the tensor forms of the above three Maxwell spinors read
\begin{align}
F^{(0)}_{ab}=2\phi_0 \bar{m}_{[a}n_{b]}+2\bar{\phi}_0m_{[a}n_{b]}\label{sec2:MaxwellField(0)},\\
F^{(1)}_{ab}=2\phi_1\left(\ell_{[a}n_{b]}+\bar{m}_{[a}m_{b]}\right)+2\bar{\phi}_1\left(\ell_{[a}n_{b]}+m_{[a}\bar{m}_{b]}\right)\label{sec2:MaxwellField(1)},\\
F^{(2)}_{ab}=2\phi_2 \ell_{[a}m_{b]}+2\bar{\phi}_2\ell_{[a}\bar{m}_{b]}\label{sec2:MaxwellField(2)}.
\end{align}}

%%%%%Setting
%\be
%\phi_0=\Phi_{AB}o^A o^B, \quad \phi_1=\Phi_{AB}o^A\iota^B,\quad \phi_2=\Phi_{AB}\iota^A\iota^B,
%\ee
%in general we have
%\be
%\Phi_{AB}=\phi_0\iota_A\iota_B-2 \phi_1 o_{(A}\iota_{B)}+\phi_2o_Ao_B.
%\ee

{\red When $n=1$, the spinor $\mc{S}$ refers to a DW spinor $\xi_A$ translated from the DW tensor}
\be\label{sec2:DW tensor}
P_{ab}=\xi_A\xi_B\varepsilon_{A'B'}.
\ee
\eq{sec2:general massless free-field equation} in this case represents the DW equation
\be\label{sec2:DW equation}
\nabla^{AA'}\xi_A=0.
\ee 
{\red The tensor form is} given by
\be\label{sec2:DW}
P_{ab}\nabla_{d}P_c^d+P_{ad}\nabla_cP_b^d=0.
\ee
{\red On the spinor dyad $(o, \iota)$, clearly, there are only two types of DW spinors: 
\begin{subequations}\label{sec2:DWspinos}
\begin{align}
\xi_A=\xi o_A\label{sec2:DWspinor-left},\\
\eta_A=\eta \iota_A\label{sec2:DWspinor-right},
\end{align}
\end{subequations}
where $\xi$ and $\eta$ are called DW scalars. Substitution of the above equations into \eq{sec2:DW equation} yields
\begin{align}
o_A \nabla^{AA'}\log\ \xi+o_A \iota^B\nabla^{AA'}o_B-\iota_A o^B \nabla^{AA'}o_B=0, \label{sec2:DWleft}\\
\iota_A \nabla^{AA'}\log\ \eta-\iota_A o^B \nabla^{AA'}\iota_B+o_A \iota^B\nabla^{AA'}\iota_B=0. \label{sec2:DWright}
\end{align}
Throughout this paper, one will find that \eq{sec2:non-degenerate Maxwell 1}-\eq{sec2:degenerate Maxwell2}, \eq{sec2:DWleft}-\eq{sec2:DWright}, and the given Bianchi identities are the basic equations for our calculation.}

It is worthwhile to mention that Dirac's equation is just a pair of coupled DW equations with {\red a} source
\be\label{sec2:Dirac equation}
\left.\begin{array}{l}
\nabla_{A^{\prime}}^{A} \xi_{A}=\mu \bar{\eta}_{A^{\prime}} \\
\nabla_{A}^{A^{\prime}} \bar{\eta}_{A^{\prime}}=\mu \xi_{A},
\end{array}\right\}
\ee
where $\mu$ is a real constant related to the mass of the spinor. The tensor version is written as 
\be
\left.\begin{array}{l}
P_{ab}\nabla_{d}P_c^d+P_{ad}\nabla_cP_b^d=-2\mu P_{ab}C_c,\\
Q_{ab}\nabla_{d}Q_c^d+Q_{ad}\nabla_cQ_b^d=-2\mu Q_{ab}C_c,
\end{array}\right\}
\ee
where $C_a=\xi_A\bar{\eta}_{\red A'}$, and the field $Q_{ab}$ written in terms of another spin-$1/2$ spinor $\eta$ reads
\be
Q_{ab}=\eta_A\eta_B\varepsilon_{A'B'}.
\ee
Although we will not use Dirac's equation in this paper, the above formulas might be useful for studying the double copy in non-vacuum spacetimes in the future.

For spin-$3/2$ massless free-fields equation, the field equation is given by 
\be\label{sec2:spin-3/2 equation}
\nabla^{AA'}\Omega_{ABC}=0.
\ee
One may refer to section 5.8 of {\red Ref.} \cite{1987ssv..book.....P} for more details about $\Omega_{ABC}$, and we will instead pay more attention to the other three massless free{\red -}fields in the following.

%%%%%%%%%%%%%%%%%%%%%%%%%%%%
\section{From gravity fields to lower spin massless free-fields}
\label{sec3:from gravity to lower spin massless free fields}

Inspired by the {\red Weyl} double copy {\blue relation} \eq{sec1:WDC}, with the fact that any massless free-field {\red spinor} with spin higher than $\frac{1}{2}$ can be constructed {\red by} scalar fields and DW {\red spinors}, we introduce a {\red general} map between vacuum gravity fields and DW fields\footnote{All of the lower spin massless free{\red-}field $({\red i}=1,2,3)$ considered in this paper  is assumed to be a test field, which will not curve the spacetime.} {\red in the curved spacetime.}
\be\label{sec3:14map}
\Psi_{ABCD}=\frac{\xi_{(A}\eta_B\zeta_C\chi_{D)}}{S_{14}}{\red .}
\ee
{\red Notably, the above} four {\red DW} spinors could be {\red identical} depending on which type of spacetime we are considering, and $S_{{\red ij}}$ is a scalar field connecting spin-${\red i}/2$  {\red spinors} {\red with} spin-${\red j}/2$ {\red spinors} e.g. {\red $i=1, j=4$} here. {\red Then,} with \eq{sec2:vacuum Einstein equation} and \eq{sec2:DW equation}, it is {\red easy} to identify what kind of {\red DW fields} can {\red exist for a specific curved spacetime}. Furthermore, if we regard {\red DW spinors} as basic units, other higher spin massless free fields {\red in} the curved spacetime are able to be constructed as well. For example, we have
\be\label{sec3:N123map}
\Phi_{AB}=\frac{\xi_{(A}\eta_{B)}}{S_{12}}
\ee
and $\Omega_{ABC}=\frac{\xi_{(A}\eta_B\zeta_{C)}}{S_{13}}$.
{\red Especially, with respect to three Maxwell spinors $\Phi^{(0)}_{AB}$, $\Phi^{(1)}_{AB}$, and $\Phi^{(2)}_{AB}$, we define the associated scalars $S_{12}$ as follows
\begin{align}\label{sec3:definition of S12}
\Phi^{(0)}_{AB}&=\frac{\eta_A\eta_B}{S^{(0)}_{12}},\quad \Phi^{(1)}_{AB}=\frac{\xi_{(A}\eta_{B)}}{S^{(1)}_{12}},\quad \Phi^{(2)}_{AB}=\frac{\xi_A\xi_B}{S^{(2)}_{12}},
\end{align}
where $\xi_A=\xi o_A$, $\eta_A=\eta\iota_A$.} {\red Based on \eq{sec3:14map} and \eq{sec3:N123map}}, it is natural to {\red lead to a} map connecting gravity fields {\red with} Maxwell fields {\red in the curved spacetime}
\be\label{sec3:24map}
\Psi_{ABCD}=\frac{\Phi_{(AB}\Theta_{CD)}}{S_{24}},
\ee
where $\Theta_{CD}$ {\red is also a Maxwell} spinor $\Theta_{CD}=\frac{\zeta_{(C}\chi_{D)}}{S'_{12}}$ with another scalar field $S'_{12}$, {\red as long as }
\be\label{sec3:general-scalar-relation124}
S_{24}=\frac{S_{14}}{S_{12}S'_{12}}.
\ee
%These are the general formula, in order to probe the physical properties of massless free-fields in the curved spacetime we still need to pay attention to exact gravity solutions.}

Notably, \eq{sec3:24map} admits a similar form {\red to the Weyl} double copy \eq{sec1:WDC}. {\red In fact,} the curved double copy for type N spacetimes and the specific relation, {\red $\Phi_{AB}=\psi^{2/3}o_{(A} \iota_{B)}$} we mentioned in Section \ref{sec:intro}, are just particular cases {\red of the present work}. {\red Along the above method,} one {\red may} ask about other situations. Is there a special relationship between different auxiliary scalar fields $S_{{\red ij}}$? What kind of spin-${\red i}/2$ massless free-field{\red s} can exist {\red in} a {\red specific} spacetime? Can we directly map a gravity field to a DW field that is living in {\red flat} space? What about mapping to Maxwell fields for type III spacetimes? We shall answer these questions in the following.

%%%%%%%%%
\subsection{Vacuum type N solutions}
\label{sec3:N}
For a vacuum type N spacetime, {\red the} Weyl tensor has only one no-vanishing component $\psi_4${\red . Combining with \eq{sec2:general Weyl},} the Weyl spinor {\red reads} $\Psi_{ABCD}=\Psi_4 o_A o_B o_C o_D$ {\red with $\Psi_4=\psi_4$}. According to \eq{sec3:14map}, there is only one {\red type} of DW spinor {\red $\xi_A$ along the basis $o$ in} the spacetime, which {\red follows} \eq{sec2:DWspinor-left}.
{\red In other words}, {\red one can always find} a special DW spinor, such that
\be\label{sec3:N14}
\Psi_{ABCD}=\frac{\xi_{(A}\xi_B\xi_C\xi_{D)}}{S_{14}}.
\ee

{\red In this case, } due to the symmetry property of $\Psi_{ABCD}$, {\red the Bianchi identity \eq{sec2:vacuum Einstein equation} are expanded into} two non{\red-}trivial dyad components
\begin{align}
o_A\nabla^{AA'}\log\ \Psi_4+4o_A \iota^B \nabla^{AA'}o_B-\iota_A o^B \nabla^{AA'}o_B=0\label{sec3:N4dyad},\\
o_A o^B \nabla^{AA'}o_B=0. 
\end{align}
Based on the Goldberg-Sachs theorem \cite{goldberg2009republication}, the congruence formed by the principal null-direction $\ell$ for algebraically specially spacetime (e.g. type N spacetime here) must be geodesic $\kappa^{\red *}=0$ and shear-free $\sigma^{\red *}=0$, {\red the} second equation {\red should hold} automatically. {\red Thus, only \eq{sec3:N4dyad} is left.} Making use of \eq{sec2:DWleft},  \eq{sec3:N14} and \eq{sec3:N4dyad}, it is not hard to identify the DW spinor $\xi_A$\footnote{{\red It is worthwhile pointing out here that \eq{sec2:DWleft} exactly verifies the statement of Ref. \cite{Godazgar:2020zbv}---the coefficient of the middle term of the left side of the equation is the rank of the corresponding spinor.}}.  Before doing that, let us {\red first} pay attention {\red to} constructing the Maxwell spinor{\red ;} then{\red ,} the problem will be {\red re}solved automatically.

{\red Since there is only one type of DW spinor $\xi_A$ in the spacetime, the unique formula of the Maxwell spinor is given by
\be
\Phi^{(2)}_{AB}=\frac{\xi^2}{S^{(2)}_{12}} o_Ao_B=\phi_2 o_A o_B\quad  \Leftrightarrow \quad \phi_2=\frac{\xi^2}{S^{(2)}_{12}}\label{sec3:Nspin1-spin1/2}.
\ee
According to \eq{sec2:degenerate Maxwell2}, one can see that there is only one independent dyad component of the field equation. Substituting \eq{sec3:Nspin1-spin1/2} into \eq{sec2:degenerate Maxwell2}, one observes that }the scalar field $S^{{\red (2)}}_{12}$ satisfies
\be\label{sec3:N12}
o_A\nabla^{AA'}\log\ S^{{\red{(2)}}}_{12}-\iota_A o^B\nabla^{AA'}o_B=0.
\ee
It can be identified by solving equations
\be\label{sec3:N12tensor}
\ell\cdot\nabla \log\ S^{{\red{(2)}}}_{12}-\rho^{\red *}=0,\quad m\cdot \nabla \log\ S^{{\red{(2)}}}_{12} - \tau^{\red *}=0.
\ee
This is an interesting result, since the scalar $S^{{\red{(2)}}}_{12}$ {\red shares} the same equation {\red with the scalar} $S_{24}$\footnote{Different from {\red the} original work, to keep consistent with the notion of this paper, here we use $S_{24}$ to denote the harmonic scalar field of {\red the} curved double copy, instead of $S$.} discovered by Ref. \cite{Godazgar:2020zbv}.
{\red Further more}, assuming spin-$3/2$ massless free-field spinor are constructed by {\red DW} spinor as follow{\red s}
\be
\Omega_{ABC}=\frac{\xi_A \xi_B \xi_C}{S_{13}}=\omega o_A o_B o_C.
\ee
Setting $S_{13}=(S^{{\red{(2)}}}_{12})^2$ (it will be soon clear why we do this),
%we will see this is reasonable soon, search for a better sentence.
combining \eq{sec2:DW equation} and \eq{sec2:spin-3/2 equation}, one will get \eq{sec3:N12} again. Therefore, for vacuum type N spacetimes, the {\red connection} between {\red Weyl spinors} and other lower spin massless free-field spinors can be summarized as {\red follows}
\be\label{sec3:Nrelationsum}
\Psi_4=\frac{\xi^4}{(S^{{\red{(2)}}}_{12})^3},\quad
\omega=\frac{\xi^3}{(S^{{\red{(2)}}}_{12})^2},\quad
\phi_2=\frac{\xi^2}{(S^{{\red{(2)}}}_{12})}.
\ee
Clearly, the curved double copy is covered by {\red the} above relations. In addition, 
{\red according to \eq{sec2:DW tensor}, the DW tensor on the null tetrad reads 
\be\label{sec3:N1tensor version}
P_{ab}=2\xi^2\ell_{[a} m_{b]}=2S^{{\red{(2)}}}_{12} \sqrt{\Psi_4 S^{{\red{(2)}}}_{12}}\ell_{[a} m_{b]}.
\ee
Using \eq{sec2:DW}, it is easy to verify whether the DW fields depend on the curvature or not.}

Next, we shall {\red show} several specific investigations {\red on} {\red exact vacuum} type N solutions.

%%%%%%%%%%%%%%%%%%%%%%
\subsubsection{Kundt solutions}
Firstly, let us focus on non-diverging solutions ($\rho^{\red *}=0$), usually, called Kundt solutions \cite{stephani2009exact}. There are two classes of Kundt solutions. One of them is plane-fronted wave with parallel propagation, called pp waves, the metric reads
\be
\mm{d} s^{2}=2 \mm{d} u(\mm{d} v+H \mm{d} u)-2 \mm{d} z \mm{d} \bar{z},
\ee
where $H(u, z, \bar{z})=f(u, z)+\bar{f}(u, \bar{z})$ with a general function $f$. Choosing {\red a} null tetrad
\be
\ell=\partial_v,\quad n=\pp_\mu- H \pp_v,\quad m=\pp_z,
\ee
one can find {\red $\tau^*=0$, and} $\Psi_4=-\pp_{\bar{z}}^2\bar{f}(u,\bar{z})$. {\red Solving \eq{sec3:N12tensor} we have} $S^{{\red (2)}}_{12}=\mc{{\red G}}(u,\bar{z})$, which is an arbitrary function of $u$ and $\bar{z}$. Therefore, {\red from \eq{sec3:Nrelationsum} the} DW scalar is {\red solved} by
\be
\xi^2=\sqrt{-\pp_{\bar{z}}^{2}\bar{f}(u,\bar{z}){\red{\mc{G}}}^{3}(u,\bar{z})}.
\ee
Clearly, due to the appearance of $\mc{{\red G}}(u,\bar{z})$, $\xi(u,\bar{z})$ can be any function of $u$ and $\bar{z}$.
{\red Moreover, t}urning to tensor version, one observes
\be
2\ell_{[a}m_{b]}=\left(\begin{array}{llll}
0\ &\ 0 &\ 0 & -1 \\
0\ &\ 0 &\ 0 &\  0 \\
0\ &\ 0 &\ 0 &\  0 \\
1\ &\ 0 &\ 0 &\  0 \\
\end{array}\right),
\ee
the only non-vanishing components of $P_{ab}$ are $P_{u\bar{z}}=-P_{\bar{z}u}=-\xi^2(u, \bar{z})$. Particularly, it is simple to check that this satisfies the DW equation {\red not only in} curved spacetime but also {\red in} Minkowski spacetime where we just need to set $H=f=0$ in the metric. 

The degenerate Maxwell {\red scalar} is then given by
\be
\phi_2=\sqrt{-\pp_{\bar{z}}^{2}\bar{f}(u,\bar{z})\mc{{\red G}}(u,\bar{z})}{\red ;}
\ee
this is nothing but the result of Ref. \cite{Godazgar:2020zbv}, which admits the Weyl double copy.

Another class is given by
\be
\begin{aligned}
\mm{d} s^{2}=2 \mm{d} u(\mm{d} v+W \mm{d} z+\bar{W} \mm{d} \bar{z}+H \mm{d} u)-2 \mm{d} z \mm{d}\bar{z},\\
W(v,z,\bar{z})=\frac{-2v}{ (z+\bar{z})},\quad H(u, v, z, \bar{z})={\red \left[f(u, z)+\bar{f}(u, \bar{z}) \right]}(z+\bar{z})-\frac{v^{2}}{(z+\bar{z})^{2}},
\end{aligned}
\ee
where $f(u, z)$ is an arbitrary function . {\red A} null tetrad is chosen as follows
\be
\ell=\partial_{v}, \quad n=\partial_{u}-(H+W \bar{W}) \partial_{v}+\bar{W} \partial_{z}+W \partial_{\bar{z}},\quad  m=\partial_{z}.
\ee
Then one obtains
\be
{\red \tau^*=-\frac{1}{z+\bar{z}},}\quad \Psi_4=-(z+\bar{z})\pp_{\bar{z}}^2\bar{f}, \quad S_{12}=\frac{\zeta(u,\bar{z})}{z+\bar{z}},
\ee
where $\zeta(u,\bar{z})$ is a{\red n arbitrary} function. So, DW scalar is given by
\be\label{sec3:N-DWin2edKundt}
\xi^2=\frac{\sqrt{-\pp_{\bar{z}}^{2}
\bar{f}(u,\bar{z})\zeta^{3}(u,\bar{z})}}{(z+\bar{z})}.
\ee
In this case
\be\label{sec3:N-DWtensorInNullIn2edKundt}
2\ell_{[a}m_{b]}=\left(\begin{array}{llll}
0\ &\ 0 &\ 0 & -1 \\
0\ &\ 0 &\ 0 &\  0 \\
0\ &\ 0 &\ 0 &\  0 \\
1\ &\ 0 &\ 0 &\  0 \\
\end{array}\right),
\ee
it is easy to check {\red that the} corresponding DW equation {\red also} holds in Minkowski spacetime. {\red As we can see from \eq{sec3:N-DWin2edKundt} and \eq{sec3:N-DWtensorInNullIn2edKundt}; the DW field is curvature-independnet}. More importantly, the degenerate Maxwell spinor is given by
\be
\phi_2=\sqrt{-\zeta(u,\bar{z})\partial^2_{\bar{z}}\bar{f}},
\ee
which is consistent with the result of Ref. \cite{Godazgar:2020zbv} and will lead to the double copy.
%%%%%%
\subsubsection{Robinson-Trautman solutions}
The solutions of general vacuum type N spacetimes admitting a geodesic, shear-free, non-twisting but diverging null congruence are given by Robinson and Trautman \cite{Robinson:1962zz,stephani2009exact}
\be
\begin{aligned}
\mm{d} s^{2}&=H \mm{d} u^{2}+2 \mm{d} u \mm{d} r-\frac{2 r^{2}}{P^{2}} \mm{d} z \mm{d} \bar{z},\\
H(u,r,z,\bar{z})&=k-2r \pp_u \log P, \quad\quad\quad (k=0,\pm1)\\
k&=2P^2 \pp_z \pp_{\bar{z}} \log P(u,z,\bar{z}).
\end{aligned}
\ee
Choosing {\red a} null tetrad as follows
\be
\ell=\partial_{r}, \quad n=\partial_{u}-\frac{1}{2} H \partial_{r}, \quad m=-\frac{P}{r} \partial_{z},
\ee
one obtains {\red $\rho^*=-1/r$, $\tau^*=0$}, and
\be
\Psi_4=\frac{P^2}{r}\pp_u\left(\frac{\pp_{\bar{z}}^2P}{P}\right),  \quad S_{12}=\frac{{\red \mc{R}}(u, \bar{z})}{r},
\ee
where $\mc{R}(u,\bar{z})$ is an arbitrary function. 
Since
\be
2P \ell_{[a}m_{b]} =\left(\begin{array}{llll}
0\ &\ 0 &\ 0 & r \\
0\ &\ 0 &\ 0 &\  0 \\
0\ &\ 0 &\ 0 &\  0 \\
-r\ &\ 0 &\ 0 &\  0 \\
\end{array}\right),
\ee
which is {\red the same} as the Kundt {\red cases;} there is only one independent component. The DW spinor is represented by
\be
\xi^2=\frac{P}{r^2}\sqrt{{\red \mc{R}}^3(u,\bar{z})\pp_u\left(\frac{\pp_{\bar{z}}^2P}{P}\right)}.
\ee 
{\red The information of the structure function $P(u, z, \bar{z})$ cannot be canceled by the function $ \mc{R}(u,\bar{z})$.} and it is easy to check that {\red the DW field} does not satisfy {\red its} field equation in Minkowski spacetime.
While for the degenerate Maxwell scalar
\be
\phi_2=\frac{P}{r}\sqrt{{\red \mc{R}}(u,\bar{z})\pp_u\left(\frac{\pp_{\bar{z}}^2P}{P}\right)},
\ee
as {\red e}xpected, it is {\blue consistent} with the result of Ref. \cite{Godazgar:2020zbv} and leads to the double copy {\blue relation}.

In summary, we rebuild the {\red Weyl} double copy with the help of DW field for non-twisting type N solutions. In addition, we find that only for the {\red Kundt class}, we can obtain a DW field such that it satisfies its field equation {\red in} Minkowski spacetime.

\subsection{Vacuum type D solutions}
\label{sec3:D}

For vacuum type D spacetimes, according to \eq{sec2:general Weyl}, the Weyl tensor has only one non-vanishing component $\psi_2$. In {\red this case}, {\red the} Weyl spinor {\red is} reduced to
\be
\Psi_{ABCD}=\Psi_2 o_{(A} o_B \iota_C \iota_{D)},
\ee
where we let $\Psi_2=6\psi_2$.
{\red The map} \eq{sec3:14map} {\red is choosen as}
\be\label{sec3:D14general}
\Psi_{ABCD}=\frac{\xi_{(A}\xi_B\eta_C\eta_{D)}}{S_{14}},
\ee
where the Weyl spinor is {\red constructed} by two mutually orthogonal DW spinors {\red given by \eq{sec2:DWspinos}
with the condition $\xi=\eta$}.
{\red Expanding the above equation on the spin bases, we obtain a} scalar identity
\be\label{sec3:D14}
\Psi_2=\frac{\xi^4}{S_{14}}.
\ee

Following the Goldberg-Sachs theorem, the congruences are formed by two principal null-directions for type D spacetimes, namely, $\ell$ and $n${\red , and} they should be geodesic and shear-free, i.e. $\kappa^{\red *}=\sigma^{\red *}=\nu^{\red *}=\lambda^{\red *}=0$. So, {\red non-trivial} dyad components of the Bianchi identity {\red \eq{sec2:vacuum Einstein equation}} are given by
\begin{align}
o_A \nabla^{AA'} \log (\Psi_2) - 3\iota_A o^B\nabla^{AA'}o_B=0,  \label{sec3:D4left}\\
\iota_A \nabla^{AA'} \log (\Psi_2) + 3o_A \iota^B\nabla^{AA'}\iota_B=0. \label{sec3:D4right}
\end{align}
In analogy to the case of type N, combining \eq {sec2:DWleft}, \eq{sec3:D14} and \eq{sec3:D4left} we have
\be\label{sec3:DS14}
o_A\nabla^{AA'}\log S_{14}+4o_A\iota^B\nabla^{AA'}o_A-\iota_A o^B\nabla^{AA'}o_B=0.
\ee
Similarly, from {\red \eq {sec2:DWright}, \eq{sec3:D14} and \eq{sec3:D4right}} we have
\be
\iota_A \nabla^{AA'}\log S_{14}-4\iota_A o^B\nabla^{AA'}\iota_B+o_A\iota^B\nabla^{AA'}\iota_B=0.
\ee
Multiplying $\bar{o}_{A'}$ and $\bar{\iota}_{A'}$ {\red on the above equations, respectively}, $S_{14}$ is solved by
\be
\begin{aligned}\label{sec3:DS14with4eqs}
\ell\cdot \nabla\log S_{14}+4\epsilon^{\red *}-\rho^{\red *}=0,\quad m\cdot\nabla\log S_{14}+4\beta^{\red *}-\tau^{\red *}=0,\\
\bar{m}\cdot\nabla\log S_{14}-4\alpha^{\red *}+\pi^{\red *}=0,\quad n\cdot \nabla\log S_{14}-4\gamma^{\red *}+\mu^{\red *}=0.
\end{aligned}
\ee
{\blue This is an overdetermined system since there is only one unknown quantity, and one will soon see that $S_{14}$ satisfies its integrability condition, so we can always find its solution. Once $S_{14}$ is solved,} {\red the DW scalars will then be identified.} {\red Different from the case of type N, since there are two types of DW spinors for vacuum type D spacetime, we can find two types of Maxwell fields in the curved spacetime. One is degenerate, the simplest forms are
\begin{align}
\Phi^{(0)}_{AB}&=\frac{\xi^2}{S^{(0)}_{12}} \iota_A\iota_B=\phi_0 \iota_A \iota_B \quad \ \Leftrightarrow \quad \phi_0=\frac{\xi^2}{S^{(0)}_{12}}
\label{sec3:D(0)spin1-spin1/2},\\
\Phi^{(2)}_{AB}&=\frac{\xi^2}{S^{(2)}_{12}} o_Ao_B=\phi_2 o_A o_B\quad\Leftrightarrow \quad \phi_2=\frac{\xi^2}{S^{(2)}_{12}}\label{sec3:D(2)spin1-spin1/2}.
\end{align}
The other one is non-degenerate, the simplest form reads
\be
\Phi^{(1)}_{AB}=\frac{\xi^2}{S^{(1)}_{12}} o_{(A}\iota_{B)}=\phi_1 o_{(A} \iota_{B)}\quad  \Leftrightarrow \quad \phi_1=\frac{\xi^2}{S^{(1)}_{12}}\label{sec3:D(1)spin1-spin1/2}.
\ee
Correspondingly, we can build two different maps between gravity fields and Maxwell fields in the curved spacetime starting from the relation \eq{sec3:D14}}

\begin{align}
\Psi_2=\frac{\frac{\xi^2}{S^{({\red 0})}_{12}}\frac{\xi^2}{S^{({\red 2})}_{12}}}{S^{(0,2)}_{24}}=\frac{\phi_{\red 0}\phi_{\red 2}}{S^{(0,2)}_{\ 24}}{\red \quad  \Leftrightarrow \quad \Psi_{ABCD}=\frac{\Phi^{(0)}_{(AB}\Phi^{(2)}_{CD)}}{S^{(0,2)}_{24}}},\label{sec3:D241} \\
\Psi_2=\frac{(\frac{\xi^2}{S^{(1)}_{12}})^2}{S^{(1,1)}_{24}}=\frac{(\phi_1)^2}{S^{(1,1)}_{\ 24}}{\red \quad  \Leftrightarrow \quad \Psi_{ABCD}=\frac{\Phi^{(1)}_{(AB}\Phi^{(1)}_{CD)}}{S^{(1,1)}_{24}}},\label{sec3:D242}
\end{align}
where upper index $(i,j)$ refers to the case of mixed {\red Maxwell scalars} $\phi_i\phi_j$. {\red One can see that the first case involves two degenerate Maxwell spinors which might lead to mixed double copy; while for} the second case, it corresponds to {\red the} classical {\red Weyl} double copy \cite{Luna:2018dpt}, {\red for which} $S^{(1,1)}_{24}=(\phi_1)^{1/2}=(\Psi_2)^{1/3}$ . We will restudy this along a new way with the help of DW {\red spinors}. {\red And we will also check whether the mixed double copy of the first case hold or not in Minkowski spacetime. The main point, in the following, is looking for source-independent Maxwell fields. }

Firstly, let us focus on degenerate Maxwell spinor{\red s}. Clearly, once DW spinors are identified, {\red to} obtain the Maxwell {\red spinor, t}he only work left for us is to identify $S_{12}$. {\red Combining \eq{sec2:degenerate Maxwell0} and \eq{sec3:D(0)spin1-spin1/2},}
$S^{(0)}_{12}$ can be solved by
\be\label{sec3:DS212tensor}
\bar{m}\cdot \nabla \log S^{(0)}_{12}+\pi^{\red *}=0,\quad n\cdot \nabla\log S^{(0)}_{12}+\mu^{\red *}=0.
\ee
{\red Analogously, $S^{(2)}_{12}$ can be solved as well; the corresponding equations have been shown in \eq{sec3:N12} for the type N case. Since the equation of $S^{(2)}_{12}$ is independent of the Petrov type of spacetime. To avoid redundancy, we will not show this equation again.}

For the second case,  substitution of \eq{sec2:DWleft} and \eq{sec3:D(1)spin1-spin1/2} into \eq{sec2:non-degenerate Maxwell 1} yields
\be
o_A\nabla\log S^{(1)}_{12}+2o_A\iota^B\nabla^{AA'}o_B=0.
\ee
Similarly, substitution of \eq{sec2:DWright} and \eq{sec3:D(1)spin1-spin1/2} into \eq{sec2:non-degenerate Maxwell 2} yields
\be
\iota_A\nabla^{AA'}\log S^{(1)}_{12}-2\iota_A o^B\nabla^{AA'}\iota_B=0.
\ee
{\red Multiplying $\bar{o}_{A'}$, $\bar{\iota}_{A'}$ respectively}, $S^{(1)}_{12}$ is solved by
\be\label{sec3:DS112}
\begin{aligned}
\ell\cdot\nabla\log S^{(1)}_{12}+2\epsilon^{\red *}=0,\quad m\cdot \nabla\log S^{(1)}_{12}+2\beta^{\red *}=0,\\
\bar{m}\cdot \nabla \log S^{(1)}_{12}-2\alpha^{\red *}=0, \quad n\cdot \nabla \log S^{(1)}_{12}-2\gamma^{\red *}=0.
\end{aligned}
\ee
{\blue This is also an overdetermined system, it is easy to check that the integrability condition holds in this case, and the solution can always be solved.}

Therefore, once $S_{14}$ is fixed, {\red we can identify the} DW field{\red s}; {\red the Maxwell fields' property then will be revealed by  $S_{12}$.} {\red If} one is instead interested in spin-$3/2$ massless free-fields, $S_{13}$ then will be the key point. {\red Thus}, it is a good starting point {\red to identify first} the DW fields {\red when investigating} the higher spin massless free-fields. Then, the rest information of the target fields will be encoded in the associated scalar fields. Once DW fields and the associated scalar fields are identified, the property of the target field{\red s should} be clear.  Using this method we can now look for source-independent electromagnetic fields for vacuum type D solutions. {\red More illustration on verifying the Weyl double copy relation will be given} with the help of {\red the} modified Plebański–Demiański {\red m}etric.

%%%%%%%%%
\subsubsection{Modified vacuum Plebański-Demiański metric}

The Plebański-Demiański metric gives a complete family of type D spacetimes \cite{PLEBANSKI197698,Griffiths:2005qp},  the original line element reads
\be
\mm{d} s^{2}=\frac{1}{(1-\hat{p} \hat{r})^{2}}\left[\frac{\mathcal{Q}\left(\mm{d} \hat{\tau}-\hat{p}^{2} \mm{d} \hat{\sigma}\right)^{2}}{\hat{r}^{2}+\hat{p}^{2}}-\frac{\mathcal{P}\left(\mm{d} \hat{\tau}+\hat{r}^{2} \mm{d} \hat{\sigma}\right)^{2}}{\hat{r}^{2}+\hat{p}^{2}}-\frac{\hat{r}^{2}+\hat{p}^{2}}{\mathcal{P}} \mm{d} \hat{p}^{2}-\frac{\hat{r}^{2}+\hat{p}^{2}}{\mathcal{Q}} \mm{d} \hat{r}^{2}\right],
\ee
where
\be
\begin{aligned}
\mathcal{P}={\red k'}+2 {\red N'} \hat{p}-{\red \epsilon'} \hat{p}^{2}+2 {\red M'} \hat{p}^{3}-\left({\red k'}+{\red e'}^{2}+{\red g'}^{2}+\Lambda / 3\right) \hat{p}^{4} \\
\mathcal{Q}=\left({\red k'}+{\red e'}^{2}+{\red g'}^{2}\right)-2 {\red M'} \hat{r}+{\red \epsilon'} \hat{r}^{2}-2 {\red N'} \hat{r}^{3}-({\red k'}+\Lambda / 3) \hat{r}^{4}.
\end{aligned}
\ee
It includes seven free real parameters, ${\red M', N', e', g', \epsilon', k',}$ and $\Lambda$. Besides the cosmological constant $\Lambda$, ${\red M'}$ is the mass parameter{\red ,} ${\red N'}$ is related to the NUT parameter, ${\red e'}$ and ${\red g'}$ are the electric and magnetic charges{\red ,} and ${\red \epsilon'}$ and ${\red k'}$ are related to the angular momentum per unit mass and the acceleration. Considering this metric cannot give {\red an obvious} physical interpretation{\red ;} for example, it is not {\red apparent} that this line element does include {\red the} well-know{\red n} Kerr metric, the NUT solution or C-metric, etc{\red ., we rescale the coordinates}
\be
\hat{p}=\sqrt{\alpha \omega} p, \quad \hat{r}=\sqrt{\frac{\alpha}{\omega}} r, \quad \hat{\sigma}=\sqrt{\frac{\omega}{\alpha^{3}}} {\red \sigma}, \quad \hat{\tau}=\sqrt{\frac{\omega}{\alpha}} {\red \tau},
\ee
and {\red the} parameters 
\be
{\red M'}+i {\red N'}=\left(\frac{\alpha}{\omega}\right)^{3 / 2}({\red M}+i {\red N}), \quad {\red \epsilon'}=\frac{\alpha}{\omega} \epsilon, \quad {\red k'}=\alpha^{2} k{\red .}
\ee
{\red A} modified metric {\red then} is given by \cite{1975AnPhy..90..196P,Griffiths:2005qp}
\be
\begin{aligned}
\mm{d} s^{2}=& \frac{1}{(1-\alpha p r)^{2}}\left[\frac{Q}{r^{2}+\omega^{2} p^{2}}\left(\mm{d} {\red \tau}-\omega p^{2} \mm{d} {\red \sigma}\right)^{2}\right.\\
&\left.-\frac{P}{r^{2}+\omega^{2} p^{2}}\left(\omega \mm{d} {\red \tau}+r^{2} d {\red \sigma}\right)^{2}-\frac{r^{2}+\omega^{2} p^{2}}{P} \mm{d} p^{2}-\frac{r^{2}+\omega^{2} p^{2}}{Q} \mm{d} r^{2}\right],
\end{aligned}
\ee
where 
\be
\begin{aligned}
P=P(p)=k+2 \omega^{-1} {\red N} p-\epsilon p^{2}+2 \alpha {\red M} p^{3}-\alpha^2\omega^2 k p^{4}, \\
Q=Q(r)=\omega^{2} k-2 {\red M} r+\epsilon r^{2}-2 \alpha \omega^{-1} {\red N} r^{3}-\alpha^{2} k r^{4}.
\end{aligned}
\ee
Since we only consider vacuum type D solutions, ${\red e', g'}$ and $\Lambda$ are set to be vanishing here. In addition, it is worthwhile to mention here that this modified metric does not include a non-singular NUT solution. In {\red practice,} to get a metric to cover all of the case{\red s}, we still need to do a coordinate transformation{\red :} $p=\frac{{\red b}}{\omega}+\frac{a}{\omega}{\red \td{p}}$, ${\red \tau}=t-\frac{(l+a)^2}{a}\phi$, and ${\red \sigma}=-\frac{\omega}{a}\phi$
where new parameters $a$ and ${\red b}$ usually correspond to a rotation parameter and a NUT parameter, respectively. However, considering the modified metric has a simple form {\red and} already covers the accelerating and rotating black hole {\red solutions} with {\red the} NUT {\red parameter}, we will use it as an example to {\red investigate the double copy} in this paper.
Choosing {\red a} null tetrad 
\be\label{sec3:type D Null tetrad}
\begin{aligned}
{\red \ell}^{\mu}=\frac{(1-\alpha p r)}{\sqrt{2\left(r^{2}+\omega^{2} p^{2}\right)}}\left[\frac{1}{\sqrt{Q}}\left(r^{2} \partial_{{\red \tau}}-\omega \partial_{{\red \sigma}}\right)-\sqrt{Q} \partial_{r}\right],\\
n^{\mu}=\frac{(1-\alpha p r)}{\sqrt{2\left(r^{2}+\omega^{2} p^{2}\right)}}\left[\frac{1}{\sqrt{Q}}\left(r^{2} \partial_{{\red \tau}}-\omega \partial_{{\red \sigma}}\right)+\sqrt{Q} \partial_{r}\right], \\
m^{\mu}=\frac{(1-\alpha p r)}{\sqrt{2\left(r^{2}+\omega^{2} p^{2}\right)}}\left[-\frac{1}{\sqrt{P}}\left(\omega p^{2} \partial_{{\red \tau}}+\partial_{{\red \sigma}}\right)+i \sqrt{P} \partial_{p}\right],
\end{aligned}
\ee
{\red we have
\be
\begin{aligned}\label{sec3:type D spin coefficients}
\rho^*=\mu^*=\frac{1+i \alpha \omega p^2}{\sqrt{2}(r+i \omega p)}\sqrt{\frac{Q(r)}{r^2+\omega^2p^2}},\\
\tau^*=\pi^*=\frac{\omega-i\alpha r^2}{\sqrt{2}(r+i\omega p)}\sqrt{\frac{P(p)}{r^2+\omega^2p^2}},\\
\epsilon^*=\gamma^*=\frac{1}{4\sqrt{2}}\left[\frac{2(1-\alpha p r)}{r+i \omega p}-2\alpha p-(1-\alpha p r)\frac{Q'}{Q}\right]\sqrt{\frac{Q(r)}{r^2+\omega^2p^2}}\\
\alpha^*=\beta^*=\frac{1}{4\sqrt{2}}\left[\frac{2\omega(1-\alpha p r)}{r+i \omega p}+2i\alpha r+i(1-\alpha p r)\frac{P'}{P}\right]\sqrt{\frac{P(p)}{r^2+\omega^2p^2}}.
\end{aligned}
\ee
\red T}he Weyl scalar is given by
\be
\Psi_2=6\psi_2=\frac{6({\red M}+i {\red N})(1-\alpha p r)^3}{(r+i \omega p)^3},
\ee
which{\blue , as we can see,} is independent of {\red coordinates} ${\red \tau}$ and ${\red \sigma}$. {\blue Plugging \eq{sec3:type D Null tetrad} and \eq{sec3:type D spin coefficients} into \eq{sec3:DS14with4eqs} we have
\be\label{sec3:type D S14}
\begin{aligned}
L_c+L_\tau\partial_\tau \log S_{14}+L_\sigma\partial_\sigma\log S_{14}+L_r\partial_r \log S_{14}=0,\\
M_c+M_\tau\partial_\tau\log S_{14}+M_\sigma\partial_\sigma\log S_{14}+M_p\partial_p\log S_{14}=0,\\
-M_c+M_\tau\partial_\tau\log S_{14}+M_\sigma\partial_\sigma\log S_{14}-M_p\partial_p\log S_{14}=0,\\
-L_c+L_\tau\partial_\tau\log S_{14}+L_\sigma\partial_\sigma \log S_{14}-L_r\partial_r\log S_{14}=0,
\end{aligned}
\ee
where
\be\begin{aligned}
L_c=\frac{Q(r)[-i+\alpha p(-3\omega p+i4r)]+(\omega p-i r)(\alpha p r-1)Q'(r)}{(\omega p-i r)\sqrt{2Q(\omega^2p^2+r^2)}},\\
L_\tau=\frac{r^2(1-\alpha p r)}{\sqrt{2Q(\omega^2p^2+r^2)}},\\
L_\sigma=-\frac{\omega(1-\alpha p r)}{\sqrt{2Q(\omega^2p^2+r^2)}},\\
L_r=-\frac{\sqrt{Q}(1-\alpha p r)}{\sqrt{2(\omega^2p^2+r^2)}},\\
M_c=\frac{P[3\alpha r^2+i\omega(4\alpha p r-1)]-i(\omega p-i r)(\alpha p r-1)P'(p)}{(\omega p -i r)\sqrt{2P(\omega^2p^2+r^2)}},\\
M_\tau=-\frac{\omega p^2(1-\alpha p r)}{\sqrt{2P(\omega^2p^2+r^2)}},\\
M_\sigma=-\frac{(1-\alpha p r)}{\sqrt{2P (\omega^2p^2+r^2)}},\\
M_p=\frac{i\sqrt{P}(1-\alpha p r)}{\sqrt{2(\omega^2p^2+r^2)}}.
\end{aligned}
\ee
From \eq{sec3:type D S14}, one can find that $S_{14}$ is independent of the coordinates $\tau$ and $\sigma$, which is the same as the Weyl scalar $\Psi_2$. The integrability condition of the above equations is then given by
\be
\partial_r\left(\frac{M_c}{M_p}\right)=\partial_p\left(\frac{L_c}{L_r}\right).
\ee
One can check that this condition does hold and} {\red we arrive at
\be
\log S_{14}=i \arctan \frac{\omega p}{r}-\log \frac{P(p)Q(r)}{\mc{C}_1(1-\alpha p r)^3 \sqrt{r^2+\omega^2 p^2}},
\ee
where $\mc{C}_1$ is an arbitrary constant of integration. Using the identity
\be\label{sec3:inverse trigonometric function}
\arctan (z)=-\frac{i}{2}\log (\frac{i-z}{i+z}) \quad \quad z\in \mathbb{C},
\ee}{\blue we obtain}
\be
S_{14}=\mc{C}_1\frac{(1-\alpha p r)^3(r+i \omega p)}{P(p)Q(r)}.
\ee
%\footnote{One may often see this symbol $\mc{C}$ in the following, we have to point out that they might not be the same one, it just denotes a constant of integration in corresponding context.}. 
{\red T}he square of {\red the} coefficient of {\red the DW} spinor $\xi_A$ reads
\be\label{sec3:the square of DW scale xi}
\xi^2=\sqrt{S_{14} \Psi_2}=\sqrt{\frac{6\mc{C}_1({\red M}+i {\red N})}{P(p)Q(r)}}\frac{(1-\alpha p r)^3}{(r+i \omega p)},
\ee
which is the coefficient of the Maxwell spinor as well as the DW {\red tensor} \eq{sec3:N1tensor version}.
It is simple to check that the DW equation {\red in the curved spacetime indeed holds in this case}. {\red Moreover}, one can see {\red that} the pre-factor $({\red M}+i {\red N})$ related to the source can be absorbed by $\mc{C}_1$, so we {\red may pay attention to} the rest term. Here, $P(p)Q(r)$ in the denominator is the only term related to the source, thus it is {\red a crutial} point when we look for a source-independent Maxwell field.

{\red On the one hand,} the degenerate Maxwell Spinor $\Phi^{({\red 0})}_{AB}$ can now be {\red identified} once scalar field $S^{({\red 0})}_{12}$ is fixed. {\red Following \eq{sec3:DS212tensor}, we arrive at}
\be
S^{({\red 0})}_{12}=\mc{C}_2\frac{1-\alpha p r}{r+i w p}\sim\frac{1-\alpha p r}{r+i \omega p},
\ee
where  $\mc{C}_2$ is an arbitrary constant of integration. {\red Analogously,} the scalar field {\red $S^{(2)}_{12}$} is shown as
\be
S^{({\red 2})}_{12}=S^{({\red 0})}_{12},
\ee
up to a constant. Notably, they are independent of the source, so $P(p)Q(r)$ term of $\xi^2$ will stay when mapped to {\red the} Maxwell field scalar,  {\red it is simple to verify that} we cannot get a source-independent degenerate Maxwell field. {\red In addition, according to \eq{sec3:general-scalar-relation124}, the scalar $S^{(0,2)}_{24}$ is given by
\be
S^{(0,2)}_{24}\sim\frac{(1-\alpha p r)(r+i \omega p)^3}{P(p)Q(r)},
\ee
one can see that it is not equal to $S^{(0)}_{12}$.}

On the other hand, for non-degenerate Maxwell spinor $\Phi^{(1)}_{AB}$, {\blue similar to the case of $S_{14}$, from} \eq{sec3:DS112} {\blue one can find that $S^{(1)}_{12}$ is independent of the coordinates $\tau$ and $\sigma$, and the integrability condition holds.} {\red With the aid of \eq{sec3:inverse trigonometric function},} we {\blue then} have
\be
S^{(1)}_{12}=\mc{C}_3\frac{(1-\alpha p r)(r+i \omega p)}{\sqrt{P(p)Q(r)}},
\ee
{\blue as one can see,} which depends on the source {\red because of the appearance of the term $P(p)Q(r)$}. {\red Furthermore}, one can see {\red that} the $P(p)Q(r)$ term of $\xi^2$ {\red in \eq{sec3:the square of DW scale xi}} will be cancel{\red l}ed out {\red by $S^{(1)}_{12}$} when mapped to $\phi_1$ {\red of \eq{sec3:D(1)spin1-spin1/2}}{\red . T}herefore, we will get a source-independent non-degenerate Maxwell field\footnote{It {\red is} easy to check that $l_{[a}n_{b]}+\bar{m}_{[a}m_{b]}$ and $l_{[a}n_{b]}+m_{[a}\bar{m}_{b]}$ are all independent of the source, from {\red the tensor form} \eq{sec2:MaxwellField(1)} one can see {\red that the} Maxwell scalar $\phi_1$ {\red is} the only physical quantity that could be affected by the source.}. {\red Besides,} $S^{(1,1)}_{24}$ is given by
\be
S^{(1,1)}_{24}=\frac{S_{14}}{(S^{(1)}_{12})^2}=\frac{\mc{C}_1}{(\mc{C}_3)^2}\frac{1-\alpha p r}{r+i \omega p}\sim\frac{1-\alpha p r}{r+i \omega p}.
\ee
Interestingly, this scalar satisfies {\red the} wave equation {\red even} in Minkowski spacetime (${\red M}={\red N}=0$). So we discover a map from a vacuum gravity field to a source-independent Maxwell field. This means that the background of the Maxwell field can be flat. One may already realize that this is nothing but {\red the Weyl} double copy.

So far, all auxiliary scalar fields connecting {\red Weyl, Maxwell} and DW fields are identified.  The maps among different spin massless-free fields are summarized as follows{\red
\be\label{sec3:D-general-maps}
\begin{aligned}
&\Psi_2=\frac{\xi^4}{(S^{(2)}_{12})^2 S^{(0,2)}_{24}}=\frac{\xi^4}{(S^{(1)}_{12})^2 S^{(1,1)}_{24}},\\ 
&S^{(0)}_{12}=S^{(2)}_{12}=S^{(1,1)}_{24}=(\phi_1)^{1/2}=(\Psi_2)^{1/3}.
\end{aligned}
\ee}%Since we have more interests in gauge field, we haven't studied the spin-$3/2$ massless free-fields here. We may discuss this in the future work. As one can see, the second line will lead to double copy. 
{\red Compared} with the case of vacuum type N {\red solutions} \eq{sec3:Nrelationsum}, $S_{24}$ is not equal to $S_{12}$ anymore {\red regarding the above two cases of the first line of \eq{sec3:D-general-maps}}. While, unexpectedly, one can see that $S^{(0)}_{12}$ and $S^{(2)}_{12}$ are equal to $S^{(1,1)}_{24}$ up to a constant. That means, the zeroth copy not only connects the vacuum gravity fields with single copy but also connects degenerate electromagnetic fields with DW fields {\red in} the curved spacetime for non-twisting {\red vacuum} type N solutions and {\red vacuum} type D solutions.
%There might exist more fundamental physics awaiting for us to probe.  All in all, Weyl double copy opens a new door for us to probe the link between gravity theory and gauge theory. 
The success of mapping gravity fields to {\red the single and zeroth copies by using the} DW spinors encourages us to extend the study to non-twisting vacuum type III solutions.

%%%%%%%%%
\subsection{Vacuum type III solutions}
For vacuum type III solutions,  $\psi_0=\psi_1=\psi_2=0$. By making a null rotation about null vector $\ell$ \cite{1998mtbh.book.....C}, 
\be\label{sec3:IIInullrotation}
\ell \to \ell, \quad n \to n+A^*m+A\bar{m}+A A^*\ell,\quad m \to m+A\ell,\quad \bar{m} \to \bar{m}+A^*\ell,
\ee
where {\red $A^*$ is the complex conjugate of a complex number $A$}, then the Weyl scalars transform like
\be
\begin{aligned}
\psi_0\to\psi_0,\quad \psi_1\to\psi_1+A^*\psi_0,\quad \psi_2\to\psi_2+2A^*\psi_1+(A^*)^2\psi_0,\\
\psi_3\to\psi_3+3A^*\psi_2+3(A^*)^2\psi_1+(A^*)^3\psi_0,\\
\psi_4\to\psi_4+4A^*\psi_3+6(A^*)^2\psi_2+4(A^*)^3\psi_1+(A^*)^4\psi_0.
\end{aligned}
\ee
Clearly, we can let $\psi_4$ vanish without changing other three Weyl scalars by requiring
\be\label{sec3:IIIAC}
A^*=-\frac{\psi_4}{4\psi_3},
\ee
and $\psi_3$ will be only non-vanishing Weyl scalar.
The spinor form thus reduces to
\be
\Psi_{ABCD}=\Psi_3o_{(A}o_B o_C \iota_{D)},
\ee
where we set $\Psi_3=-4\psi_3$.
Based on this, \eq{sec3:14map} {\red can be written in the form}
\be\label{sec3:III13general}
\Psi_{ABCD}=\frac{\xi_{(A}\xi_B\xi_C\eta_{D)}}{S_{14}},
\ee
where $\xi_A=\xi o_A$ {\red and} $\eta_A=\eta \iota_A$, or in scalar from
\be\label{sec3:III13generalscalar}
\Psi_3=\frac{\xi^3\eta}{S_{14}}.
\ee

{\red According to \eq{sec2:vacuum Einstein equation}, t}wo independent dyad components of the Bianchi identity read
\begin{align}
o_A \nabla^{AA'} \log \Psi_3 +2\nabla^{AA'}o_A=0,  \label{sec3:III4left}\\
\iota_A \nabla^{AA'} \log \Psi_3 + 4o_A \iota^B\nabla^{AA'}\iota_B+2\iota_Ao^B\nabla^{AA'}\iota_B=0. \label{sec3:III4right}
\end{align}
{\red Since
\be
\begin{aligned}
o_A \iota^B\nabla^{AA'}o_B-\iota_A o^B \nabla^{AA'}o_B&=(o_A \iota^B-\iota_A o^B)\nabla^{AA'}o_B\\
                                                                                          &=\epsilon_A^{\ B}\nabla^{AA'}o_B\\
                                                                                          &=\nabla^{AA'}(\epsilon_A^{\ B}o_B)\\
                                                                                          &=\nabla^{AA'}o_A,
\end{aligned}
\ee
\eq{sec2:DWleft} can be rewritten as
\be
o_A \nabla^{AA'} \log \xi +\nabla^{AA'}o_A=0\label{sec3:DWnewform}.
\ee
Combining} \eq{sec3:III4left} and \eq{sec3:DWnewform}, we {\red have}
\be\label{sec3:III xi}
\Psi_3=\mc{C}\xi^2,
\ee
where $\mc{C}$ is a {\red non-vanishing} constant of integration. This result is even independent of our assumption \eq{sec3:III13general}. {\red I}n order to keep {\red the} total spin invariant for the above equation\footnote{We thank Ricardo Monteiro for bringing this up.}, the constant $\mc{C}$ here should {\red correspond} to a field with {\red a} total spin of $1$. {\red However}, we need to point out that it is not necessary to require it to be a source-free Maxwell scalar. This point can also be verified from \eq{sec3:III13general} or \eq{sec3:III13generalscalar}, then one observes
\be\label{sec3:IIIC}
\mc{C}=\frac{\xi\eta}{S_{14}}{\red ;}
\ee
{\red t}his is nothing but a constraint equation about spinor $\xi_A$ and spinor $\eta_A$. Furthermore, one can see {\red that} $\mc{C}$ indeed corresponds to a field with {\red a} total spin of $1$ in view of the right side of \eq{sec3:IIIC}. {\red Yet} there is no reason to require $S_{14}=S_{12}$, $\mc{C}$ does not have to be a source-free Maxwell scalar. We will come back to talk more about this in the discussions. {\red Regarding spinor $\xi_A$, according to \eq{sec3:definition of S12}, we can construct a degenerate Maxwell spinor $\Phi^{(2)}_{AB}$ with it. Repeating the same calculation as the case of type N,  $S^{(2)}_{12}$ can be solved, and then the Maxwell field will be identified.}

As for another DW spinor $\eta_A=\eta \iota_A$, the equation of motion is given by 
\be
\iota_A\nabla^{AA'}\log (\frac{\Psi_3}{\eta})+3o_A\iota^B\nabla^{AA'}\iota_B+3\iota_A o^B\nabla^{AA'}\iota_B=0
\ee
following \eq{sec2:DWright}  and \eq{sec3:III4right}.
The tensor version then reads
\begin{align}\label{sec3:III eta}
 \bar{m} \cdot \nabla \log \left(\frac{\Psi_3}{\eta}\right)+3\pi^{\red *} +3 \alpha^{\red *}=0, \quad
n \cdot \nabla \log \left(\frac{\Psi_3}{\eta}\right)+3\mu^{\red *} +3 \gamma^{\red *}=0.
\end{align}
{\blue Recalling the type N and type D cases, the DW scalars mapped from the gravity fields all depend on the same coordinates as the Weyl scalars, we expect $\eta$ also behaves like that and we are in fact only interested in this case in the present work. However, one will see that its solution is also related to the other coordinates unless we impose an extra condition. Therefore, generally, there is no trivial relationship between the Weyl scalar $\Psi_3$ and the DW scalar $\eta$, we thus shall pay more attention to the DW tensor $\xi_A$ in this section.}

{\red Further investigation on exact} non-twisting vacuum type III {\red solutions is given in the following}.

%%%%%%%%%%%%%%
\subsubsection{Kundt solutions}
{\red There are two kinds of Kundt solutions for the type III case, the metric in general }is given by \cite{stephani2009exact}
\be
\mm{d}s^2=2{\red \mm{d}u (H\mm{d}u+\mm{d}v+W\mm{d}z+\bar{W}\mm{d}\bar{z})}-2\mm{d}z \mm{d}\bar{z},
\ee
{\red with a real function $H$ and a complex function $W$.}

{\red For the case of $W_{,v}=0$, }
\be
\begin{aligned}
W=W(u,\bar{z}),\quad H=\frac{1}{2}\left(W_{,\bar{z}}+\bar{W}_{,z}\right)v+H^0,\\
H^0_{,z\bar{z}}-\mathfrak{Re}\left[W^2_{,\bar{z}}+W W_{,z\bar{z}}+W_{,u\bar{z}}\right]=0.
\end{aligned}
\ee
{\red We choose a} null tetrad
\begin{align}
\ell=\partial_v, \quad n=\partial_u-(H+W\bar{W})\partial_v+\bar{W}\partial_z+W\partial_{\bar{z}}, \quad m=\partial_{\bar{z}}.
\end{align}
The Weyl scalars {\red in this case} are given by
\be
\psi'_3={\red -}\ \frac{1}{2}\partial^2_{\bar{z}}W(u,\bar{z}),\quad \psi'_4={
\red -}\ \bar{W}(u,z)\partial^2_{\bar{z}}{\red W}(u,\bar{z})\ {\red -}\ \frac{1}{2}v\partial^3_{\bar{z}}W(u,\bar{z})\ {\red -}\ \partial^2_{\bar{z}}H^0(u,z,\bar{z}).
\ee
Note $\partial^2_{\bar{z}}W(u,\bar{z})\neq0$ here, otherwise the metic reduces to type N solution. By making a null rotation with the help of \eq{sec3:IIIAC}, the only non-vanishing Weyl scalar left is \cite{Pravda:2002us}
\be
\Psi_3=-4\psi_3\ {\red =}\ 2\partial^2_{\bar{z}}W(u,\bar{z}).
\ee
From \eq{sec3:III xi}, one of the {\red corresponding} Dirac-Weyl fields is given by
\be
\xi^2{\red =}\frac{2}{\mc{C}}\partial^2_{\bar{z}}W(u,\bar{z}).
\ee
In addition, {\blue some spin coefficients are given by
\be
\begin{aligned} 
\rho^*=\tau^*=\alpha^*=0,\\
\mu^*=\frac{[\pp^3_{\bar{z}}W(2W\pp^2_z\bar{W}+v\pp^3_z\bar{W}+2\pp^2_{\bar{z}}H^0)+8\pp^2_z\bar{W}(\pp_{\bar{z}}W\pp^2_{\bar{z}}W-\pp_z\pp^2_{\bar{z}}H^0)]}{16\pp^2_{\bar{z}}W\pp^2_z\bar{W}},\\
\pi^*=-\frac{\pp^3_{\bar{z}}W}{4\pp^2_{\bar{z}}W},\\
\gamma^*=\frac{1}{2}\pp_{\bar{z}}W.
\end{aligned}
\ee
} solving \eq{sec3:N12tensor} {\red we have $\partial_v S^{(2)}_{12}=\partial_zS^{(2)}_{12}=0$.} $S^{(2)}_{12}$ is independent of $v$ and $z$, {\red namely}, it can be an arbitrary function of $u$ and $\bar{z}$,
\be
S^{(2)}_{12}=S^{(2)}_{12}(u,\bar{z}).
\ee
{\red And, it is easy to check that $S^{(2)}_{12}$ satisfies the wave equation even in the flat spacetime.}
Then the degenerate Maxwell scalar is given by
\be
\phi_2=\frac{\xi^2}{S^{(2)}_{12}}{\red =}\frac{2}{\mc{C}}\frac{\partial^2_{\bar{z}}W(u,\bar{z})}{S^{(2)}_{12}(u,\bar{z})}.
\ee
Combining the fact
\be
2\ell_{[a}m_{b]}=\left(\begin{array}{llll}
0\ &\ 0 &\ 0 & -1 \\
0\ &\ 0 &\ 0 &\  0 \\
0\ &\ 0 &\ 0 &\  0 \\
1\ &\ 0 &\ 0 &\  0 \\
\end{array}\right)\quad\text{and}\quad 
2\ell_{[a}\bar{m}_{b]}=\left(\begin{array}{llll}
0\ &\ 0 &-1 &\ 0 \\
0\ &\ 0 &\ 0 &\ 0 \\
1\ &\ 0 &\ 0 &\  0 \\
0\ &\ 0 &\ 0 &\  0 \\
\end{array}\right),
\ee
{\red one can find} that this degenerate Maxwell field {\red also} satisfies the field equation {\red in} Minkowski spacetime, {\red for which} we {\red may} let $W=H^{\red 0}=0$ {\red in the metric}. Recalling the relationship \eq{sec3:III xi}, the Weyl scalar can be written as
\be\label{sec3:IIIKunt24}
\Psi_3= \mc{C}S^{(2)}_{12} \phi_2{\red .}
\ee

{\red Besides}, we also {\red probe} another {\red DW spinor's} form. {\blue Keeping consistent with the Weyl scalar $\Psi_3$, we are only interested in the solution that $\eta$ is independent of the coordinates $v$ and $z$. In this case, by solving \eq{sec3:III eta} we obtain
\be 
\pp_u\log (\frac{\Psi_3}{\eta})=\mc{M},\quad \pp_{\bar{z}}\log (\frac{\Psi_3}{\eta})=\mc{N},
\ee
where
\be
\begin{aligned}
\mc{M}=-\frac{3(4\pp_{\bar{z}}W\pp^2_{\bar{z}}W+W\pp^3_{\bar{z}}W-2\pp_z\pp^2_{\bar{z}}H^0)}{4\pp^2_{\bar{z}}W},\\
\mc{N}=\frac{3\pp^3_{\bar{z}}W}{4\pp^2_{\bar{z}}W}.
\end{aligned}
\ee
The integrability condition is given by $\pp_{\bar{z}}\mc{M}=\pp_u \mc{N}$. Clearly, to have a solution we have to impose one more condition, $\pp^2_z\pp^2_{\bar{z}}H^0=0$. In general, however, there is no solution which depends on the same coordinates as the Weyl scalar, and} there is no trivial relation between DW scalar $\eta_A$ and Weyl scalar $\Psi_3$. {\blue In the following, we will focus on the DW tensor $\xi_A$.}

{\red For the case of} $W_{,v}\ne 0${\red ,}
\be
\begin{aligned}
W=&W^0(u,z)-\frac{2v}{z+\bar{z}}, \quad H=H^0+v\frac{W^0+\bar{W}^0}{z+\bar{z}}-\frac{v^2}{(z+\bar{z}){\red ^2}},\\
&\left(\frac{H^0+W^0\bar{W}^0}{z+\bar{z}}\right)_{,z\bar{z}}=\frac{W^0_{,z}\bar{W}^0_{,\bar{z}}}{z+\bar{z}}.
\end{aligned}
\ee
{\red We choose a null tetrad
\begin{align}
\ell=\partial_v, \quad n=\partial_u-(H+W\bar{W})\partial_v+\bar{W}\partial_z+W\partial_{\bar{z}}, \quad m=\partial_{\bar{z}}.
\end{align}}
By doing a null rotation {\red with} \eq{sec3:IIIAC}, the Weyl scalar is given by \cite{Pravda:2002us}
\be
\Psi_3={\red -4\psi_3=4}\frac{\partial_{\bar{z}}\bar{W}^0(u,\bar{z})}{z+\bar{z}}.
\ee
Correspondingly, {\red we arrive at}
\be
\xi^2=\frac{1}{\mc{C}}\Psi_3\ {\red =}\ \frac{4}{\mc{C}}\frac{\partial_{\bar{z}}\bar{W}^0(u,\bar{z})}{z+\bar{z}}.
\ee
{\red The spin coefficients $\rho^*$ and $\tau^*$ are given by
\be
\rho^*=0, \quad \tau^*=-\frac{1}{z+\bar{z}}.
\ee}
{\red Following \eq{sec3:N12tensor}, the} auxiliary scalar field {\red is solved by}
\be
S^{(2)}_{12}=\frac{{\red \mc{V}}(u,\bar{z})}{z+\bar{z}}{\red ,}
\ee
{\red where function $\mc{V}(u,\bar{z})$ is arbitrary. One can check that $S^{(2)}_{12}$ satisfies the wave equation even in the flat spaceitme.
Moreover}, we have
\be
\phi_2{\red =}\frac{4}{\mc{C}}\frac{\partial_{\bar{z}}\bar{W}^0(u,\bar{z})}{{\red \mc{V}}(u,\bar{z})}.
\ee
Since 
\be
2\ell_{[a}m_{b]}=\left(\begin{array}{llll}
0\ &\ 0 &\ 0 & -1 \\
0\ &\ 0 &\ 0 &\  0 \\
0\ &\ 0 &\ 0 &\  0 \\
1\ &\ 0 &\ 0 &\  0 \\
\end{array}\right),\quad
2\ell_{[a}\bar{m}_{b]}=\left(\begin{array}{llll}
0\ &\ 0 &-1 &\ 0 \\
0\ &\ 0 &\ 0 &\ 0 \\
1\ &\ 0 &\ 0 &\  0 \\
0\ &\ 0 &\ 0 &\  0 \\
\end{array}\right),
\ee
similar to the case {\red of} $W_{,v}=0$, one can show {\red that} the Maxwell field also satisfies its field equation {\red in} Minkowski space. The Weyl scalar {\red is} written as 
\be
\Psi_3=\mc{C} S^{(2)}_{12}\phi_2.
\ee

%%%%%%%%%%%%%
\subsubsection{Robinson-Trautman solutions}

The vacuum solution of diverging non-twisting {\red type III} case is given by \cite{Robinson:1962zz,stephani2009exact},
\be
\begin{aligned}
\mm{d}s^2={\red \mm{d}u(H \mm{d}u+2\mm{d}r)}-\frac{2r^2}{P^2(u,z,\bar{z})}\mm{d}z \mm{d}\bar{z},\\
\Delta\log P=\mc{K}=-3\left[f(u,z)+\bar{f}(u,\bar{z})\right],\quad  f_{,z}\ne 0,\\
H=\Delta \log P-2r\partial_u \log P,\quad \Delta \equiv 2P^2\partial_z\partial_{\bar{z}}.
\end{aligned}
\ee
where {\red the structure function f(u,z) is complex}.

Choosing {\red a} null tetrad
\be
\ell=\partial_r,\quad n=\partial_u-\frac{H}{2} \partial_r,\quad m=-\frac{P}{r}\partial_z,
\ee
the non-vanishing Weyl scalars {\red read}
\be
\begin{aligned}
\psi_3'&{\red =}\frac{3P\partial_{\bar{z}}\bar{f}}{2r^2}, \\
 \psi_4'&=\frac{3P^2\partial^2_{\bar{z}}\bar{f}{\red -}2r \partial^2_{\bar{z}}P\partial_u P{\red +}2P(3\partial_{\bar{z}}P\partial_{\bar{z}}\bar{f}+r\partial_u\partial^2_{\bar{z}}P)}{2r^2}.
\end{aligned}
\ee
Same to the case of Kundt class {\red of the} last section, by doing a null rotation {\red with} \eq{sec3:IIIAC}, the only non-vanishing Weyl scalar {\red reads} 
\be
\Psi_3=-4\psi_3=-4\psi_3'={\red -}\frac{6P\partial_{\bar{z}}\bar{f}}{r^2}
\ee 
In the new null tetrad, according to \eq{sec3:III xi}, one of the DW fields mapping from {\red the} gravity side {\red is given by}
\be
\xi^2={\red -}\frac{6}{\mc{C}}\frac{P\partial_{\bar{z}}\bar{f}}{r^2}.
\ee
{\red The spin coefficients $\rho^*$ and $\tau^*$ are solved by
\be
\begin{aligned}
\rho^*=-\frac{1}{r},\\
\tau^*=\frac{3P^2\pp^2_{z} f-2r\pp_u P\pp^2_{\bar{z}}P+2P(3\pp_z f \pp_{\bar{z}}P+r\pp_u\pp^2_{\bar{z}}P)}{12P r\pp_z f}.
\end{aligned}
\ee}
Making use of \eq{sec3:N12tensor}, we find $S^{(2)}_{12}$ has to satisfy
\be
\frac{1}{r}+{\frac{\partial_r S^{(2)}_{12}}{{\red S^{(2)}_{12}}}}=0,\quad \partial_z S^{(2)}_{12}=0.
\ee
Therefore, we arrive at a general solution
\be
S^{(2)}_{12}=\frac{\mc{X}(u,\bar{z})}{r},
\ee
where $\mc{X}(u, {\red \bar{z}})$ is an arbitrary function. {\red F}rom \eq{sec3:D(2)spin1-spin1/2} the degenerate Maxwell scalar reads
\be\label{sec3:IIIRT212}
\phi_2=\frac{\xi^2}{S^{(2)}_{12}}={\red -}\frac{6}{\mc{C}}\frac{P\partial_{\bar{z}}\bar{f}(u,\bar{z})}{r \mc{X}(u,\bar{z})}.
\ee
Going to {\red the} tensor version \eq{sec2:MaxwellField(2)}, we have
\be
2\frac{P}{r}\ell_{[a}m_{b]}=\left(\begin{array}{llll}
\ 0\ &\ 0 &\ 0 &\ 1 \\
\ 0\ &\ 0 &\ 0 &\  0 \\
\ 0\ &\ 0 &\ 0 &\  0 \\
-1\ &\ 0 &\ 0 &\  0 \\
\end{array}\right),\quad 2\frac{P}{r}\ell_{[a}\bar{m}_{b]}=\left(\begin{array}{llll}
\ 0\ &\ 0 &\ 1 &\ 0 \\
\ 0\ &\ 0 &\ 0 &\  0 \\
-1 \ &\ 0 &\ 0 &\  0 \\
\ 0\ &\ 0 &\ 0 &\  0 \\
\end{array}\right).
\ee
Clearly, only the $[u z]$ and $[u \bar{z}]$ components are non-vanishing{\red . S}imilar to the case of type N, it is easy to show that this field satisfies {\red the} field equation {\red in} Minkowski spacetime. {\red In addition, c}ombining \eq{sec3:III xi} and \eq{sec3:IIIRT212}, we have
\be\label{sec3:III24general}
\Psi_3=\mc{C} S^{(2)}_{12}\phi_2,
\ee
where the scalar field $S^{(2)}_{12}$ satisfies {\red the} wave equation not only {\red in} this curved spacetime but also {\red in} Minkowski spacetime. 

Therefore, with the help of the DW {\red spinor}s, we have successfully proved that there indeed exists a natural map between
pure Maxwell fields and gravity fields for non-twisting vacuum type III spacetimes. {\red Moreover, we found that} the auxiliary scalar field, {\red connecting} the DW field {\red with the degenerate electromagnetic field in the curved spacetime}, plays a similar role to the zeroth copy.  
%%%%%%%%%%%%%%%%%%%%%%%%%%%%
\section{Discussion and Conclusions}
\label{sec4:discussion and conclusion}
In this paper, based on the fact that any massless free-field {\red spinors} with spin higher than {\red$1/2$} can be constructed {\red with} {\red DW spinors} (spin-{\red $1/2$}) and scalar fields, we introduced a map between vacuum gravity fields and DW fields in spin-space{\red . T}he form of associated DW spinors are identified. Regarding these DW spinors as basic units, we investigated the other higher spin massless-free fields{\red , especially the Maxwell fields,} and showed some hidden fundamental features among these fields. 

{\red In particular}, for Petrov type N {\red solutions}, inspired by the work \cite{Godazgar:2020zbv}, we found {\red that} only one type of DW spinor {\red exists in the curved spacetime;} {\red combining with the zeroth copy, the DW spinor can construct any other higher spin massless free-fields.} Following this, we studied the Petrov type D {\red solution}s. In this situation, there are two types of DW spinors {\red in the curved spacetime}. {\red Unlike} the case of type N, we found {\red that} $S_{24}$ {\red (a scalar field connecting a Maxwell field with a gravity field)} is not equal to $S_{12}$ {\red (a scalar field connecting a DW field with a Maxwell field)} an{\red ym}ore for each case. While, there remains an interesting relation{\red , $S^{(0)}_{12}=S^{(2)}_{12}=S^{(1,1)}_{24}$,} the scalar fields {\red connecting the DW fields with the} degenerate electromagnetic fields are equal to the zeroth copy up to a constant. In general, by {\red using} the DW {\red spinors} and the auxiliary scalar fields, we systematically rebuilt the Weyl double copy for non-twisting vacuum type N and vacuum type D {\red solution}s in this paper. Our result{\red s} are consistent with previous work {\red \cite{Luna:2018dpt,Godazgar:2020zbv}}. {\red Moreover}, we showed that the zeroth copy not only connects the gravity fields with {\red the} single copy but also connects DW fields with {\blue those degenerate electromagnetic fields} living in the curved spacetime. 

We also investigated the case of {\red non-twisting} vacuum type III solutions{\red . I}ndependent of the {\red proposed} map, we found that the square of a DW scalar {\red is} just proportional to the Weyl scalar $\Psi_3$. {\red Such an} interesting result produces a natural relationship between {\red the} gravity fields and {\red the} Maxwell fields {\red in the flat spacetime}, which {\red is} summarized as $\Psi_3=\mc{C}S^{(2)}_{12}\phi_2$, {\blue where} $S^{(2)}_{12}$ and $\phi_2$ correspond to a scalar field and a degenerate Maxwell field{\red ,} respectively. Interestingly, both of them not only satisfy their field equation {\red in} curved spacetime but also {\red in} Minkowski spacetime. {\red A}s an auxiliary scalar field associated {\red with the} degenerate electromagnetic field, it is not surprising that $S^{(2)}_{12}$ plays a role similar to the zeroth copy {\red considering} our discovery in the cases of type N and type D solutions. However{\red ,} why this scalar can play such {\red an} important role {\red in} connecting gravity {\red theory} with gauge theory is still unclear.  {\red On the whole}, with the help of the chosen DW spinors, {\red we} systematically show that there indeed exists a deep connection between gravity theory and gauge theory by investigating non-twisting vacuum type N, III and vacuum type D solutions. {\red T}he Weyl double copy {\red proposed before} is covered {\red in the present work}.

Next, it would be fascinating to study the {\red case in} non-vacuum {\red spacetime using} Dirac equation \eq{sec2:Dirac equation} instead of Dirac-Weyl equation \eq{sec2:DW equation}. {\red T}he situation {\red could} be viewed as turning from a DW equation to DW equations with {\red a} source. In addition, so far, all of the works related to the Weyl double copy only focus on classical gravity solutions without a cosmological constant{\red . Along the road of this work, i}t would be interesting to show {\red a} specific situation about the {\red Weyl} double copy for asymptotically {\red (anti-)}de Sitter spacetimes. In fact, we found that the Weyl double copy{\red ,} in general{\red ,} satisf{\red ies} conformally invariant field equations {\red even in} conformally flat space{\red times}, which is consistent with the result of twistorial version of Weyl double copy\cite{White:2020sfn}. Progress on this {\red has been shown} in another work \cite{Han:2022mze}.

In the end, we have to point out {\red that} although we have shown a natural map for type III cases between gravity fields and the Maxwell fields living in Minkowski spacetime, we did not prove if type III spacetime admits the {\red classical Weyl} double copy {\blue prescription}. {\red In terms of} Kundt class {\blue with $W_{,v}=0$}, we only shown {\red that} the DW scalar $\eta$ {\blue does not depends on the same coordinates as the Weyl scalar, unless we impose one more condition --- $H^0_{,zz\bar{z}\bar{z}}=0$.} {\red If the Weyl double copy prescription does exist} for vacuum type III solutions, the {\red possible} way to {\red show} it may start from regarding $S^{(2)}_{12}$ as {\red the} zeroth copy{\red .} {\red Then, it would be interesting to probe} the physical meaning of the constant $\mc{C}$, since it corresponds to a field with {\red a} total spin of $1$. {\red Alternatively}, we may need to extend the Weyl double copy to a more general form to cover even the twisting case. All in all, to get full knowledge about the relation between gravity theory and gauge theory, there is still a long way to go. We hope this paper provides new insights for a better understanding of double copy and the connection between gravity theory and gauge theory.

\section*{Acknowledgements}
The author would like to thank Andr\'es Luna, Niels A. Obers and Xin Qian for helpful discussions. The author also would like to thank Ricardo Monteiro and Niels A. Obers for instructive comments on the manuscript. The author thanks the Theoretical Particle Physics and Cosmology section at the Niels Bohr Institute for support. This work is also financially supported by the China Scholarship Council.

\bibliographystyle{JHEP}
\bibliography{refss}

\end{document}